\documentclass{aa}  
\usepackage{graphicx,upgreek,txfonts,aas_macros}
\usepackage{multirow}   

\usepackage[hidelinks]{hyperref}
\usepackage{color}
\begin{document} 

\def\xmm {\emph{XMM--Newton}}
\def\cxo {\emph{Chandra}}
\def\swift {\emph{Swift}}
\def\frm {\emph{Fermi}}
\def\igr {\emph{INTEGRAL}}
\def\sax {\emph{BeppoSAX}}
\def\xte {\emph{RXTE}}
\def\rst {\emph{ROSAT}}
\def\asca {\emph{ASCA}}
\def\hst {\emph{HST}}
\def\nst {\emph{NuSTAR}}
\def\gaia {\emph{Gaia}}
\def\srclong {\mbox{CXOU\,J005440.5--374320}}
\def\src {\mbox{J0054}}
\def\flux {\mbox{erg cm$^{-2}$ s$^{-1}$}}
\def\lum {\mbox{erg s$^{-1}$}}
\def\nh {$N_{\rm H}$}
\newcommand{\rev}[1]{{ #1}}
\newcommand{\revv}[1]{{ #1}}
\newcommand{\revvv}[1]{{ #1}}
\newcommand{\revvvv}[1]{{ #1}}
\newcommand{\revV}[1]{{ #1}}


   \title{A soft and transient ultraluminous X-ray source with 6-h modulation in the NGC\,300 galaxy}

\titlerunning{A transient ULX in NGC\,300 with 6-h modulation}
\authorrunning{A. Sacchi et al.}

   \author{A.\,Sacchi,\inst{1,2} 
   P.\,Esposito,\inst{1,3}  
   D.\,de\,Martino,\inst{4}
   R.\,Soria,\inst{5,6,7}
   G.\,L.\,Israel,\inst{8} 
   A.\,A.\,C. Sander,\inst{9} 
   L.\,Sidoli,\inst{3} 
   D.\,A.\,H.\,Buckley,\inst{10,11,12}\\ 
   I.\,M.\,Monageng,\inst{10,11}
   A.\,Tiengo,\inst{1,3}
   M.\,Arca\,Sedda,\inst{13,14,9}
   C.\,Pinto,\inst{15}
   R.\,Di\,Stefano,\inst{2}
   M.\,Imbrogno,\inst{16,8}
   A.\,Carleo,\inst{17,18}
   G.\,Rivolta\inst{19}
    }

\institute{Scuola Universitaria Superiore IUSS Pavia, Palazzo del Broletto, piazza della Vittoria 15, I-27100 Pavia, Italy\\
e-mail: \href{mailto:andrea.sacchi@cfa.harvard.edu}{andrea.sacchi@cfa.harvard.edu}, \href{mailto:paolo.esposito@iusspavia.it}{paolo.esposito@iusspavia.it}
\and Center for Astrophysics $\vert$ Harvard \& Smithsonian, 60 Garden Street, Cambridge, MA 02138, USA
\and INAF--Istituto di Astrofisica Spaziale e Fisica Cosmica di Milano, via A. Corti 12, I-20133 Milano, Italy
\and INAF--Osservatorio Astronomico di Capodimonte, salita Moiariello 16, I-80131, Napoli, Italy
\and College of Astronomy and Space Sciences, University of the Chinese Academy of Sciences, Beijing 100049, China
\and INAF--Osservatorio Astrofisico di Torino, Strada Osservatorio 20, I-10025 Pino Torinese, Italy
\and Sydney Institute for Astronomy, School of Physics A28, The University of Sydney, Sydney, NSW 2006, Australia
\and INAF--Osservatorio Astronomico di Roma, via Frascati 33, I-00078 Monteporzio Catone, Italy
\and Zentrum für Astronomie der Universit\"at Heidelberg, Astronomisches Rechen-Institut, Mönchhofstr. 12-14, D-69120 Heidelberg,
Germany
\and South African Astronomical Observatory, P.O. Box 9, 7935 Observatory, South Africa
\and Department of Astronomy, University of Cape Town, Private Bag X3, 7701 Rondebosch, South Africa
\and Department of Physics, University of the Free State, PO Box 339, Bloemfontein 9300, South Africa
\and Gran Sasso Science Institute (GSSI), viale Francesco Crispi 7, I-67100 L’Aquila, Italy
\and Physics and Astronomy Department Galileo Galilei, University of Padova, Vicolo dell’Osservatorio 3, I-35122, Padova, Italy
\and INAF--Istituto di Astrofisica Spaziale e Fisica Cosmica di Palermo, Via U. La Malfa 153, I-90146 Palermo, Italy
\and Dipartimento di Fisica, Università degli Studi di Roma ``Tor Vergata'', via della Ricerca Scientifica 1, I-00133 Roma, Italy
\and Dipartimento di Fisica  ``E.R. Caianiello'', Universit\'a di Salerno and INFN Sezione di Napoli (Gruppo Collegato di Salerno), via Giovanni Paolo II 132, I-84084 Fisciano, Italy
\and INAF--Osservatorio Astronomico di Cagliari, via della Scienza 5, I-09047 Selargius, Italy
\and Dipartimento di Fisica ``Aldo Pontremoli'', Universit\`a degli Studi di Milano, via G. Celoria 16, I-20133 Milano, Italy
}
              
  \date{Received DD Month YYYY; accepted DD Month YYYY}

  \abstract{We investigate the nature of CXOU\,J005440.5-374320 (\src), a peculiar bright ($\sim$$4\times10^{39}$\,\lum) and soft X-ray transient in the spiral galaxy NGC\,300 with a 6-hour periodic flux modulation that was detected in a 2014 \cxo\ observation.  Subsequent observations with \cxo\ and \xmm, as well as a large observational campaign of NGC\,300 and its sources performed with the \emph{Swift Neil Gehrels Observatory}, showed that this source exhibits recurrent flaring activity: four other outbursts were detected across $\sim$8 years of monitoring. Using data from the \swift/UVOT archive and from the \xmm/OM and \gaia\ catalogues, we noted the source is likely associated with a bright blue optical/ultraviolet counterpart. This prompted us to perform \rev{follow-up observations} with the \emph{Southern African Large Telescope} in December 2019. With the multi-wavelength information at hand, we discuss several possibilities for the nature of \src. 
   Although none is able to account for the full range of the observed peculiar features, we found that the two most promising scenarios are a stellar-mass \rev{compact object} in a binary system with a Wolf--Rayet star companion, or the recurrent tidal stripping of a stellar object trapped in a system with an intermediate-mass ($\sim$1000\,M$_\odot$) black hole.}
  \keywords{galaxies:individual: NGC\,300 -- X-rays:individuals: CXOU\,J005440.5--374320}
  \maketitle

\section{Introduction}

The \cxo\ ACIS Timing Survey at Brera And Roma astronomical observatories (CATS\,@\,BAR) is a systematic search for coherent periodic signals in \cxo/ACIS archival data in Timed Exposure mode. The first catalog of new pulsators was published in \citet{israel16} and several sources were studied individually \citep{eis13,eisrc13,eism13,eimm15,bartlett17,sidoli16,sidoli17}. Over 50 new X-ray pulsators were discovered so far.

\begin{figure*}
	\includegraphics[width=\hsize]{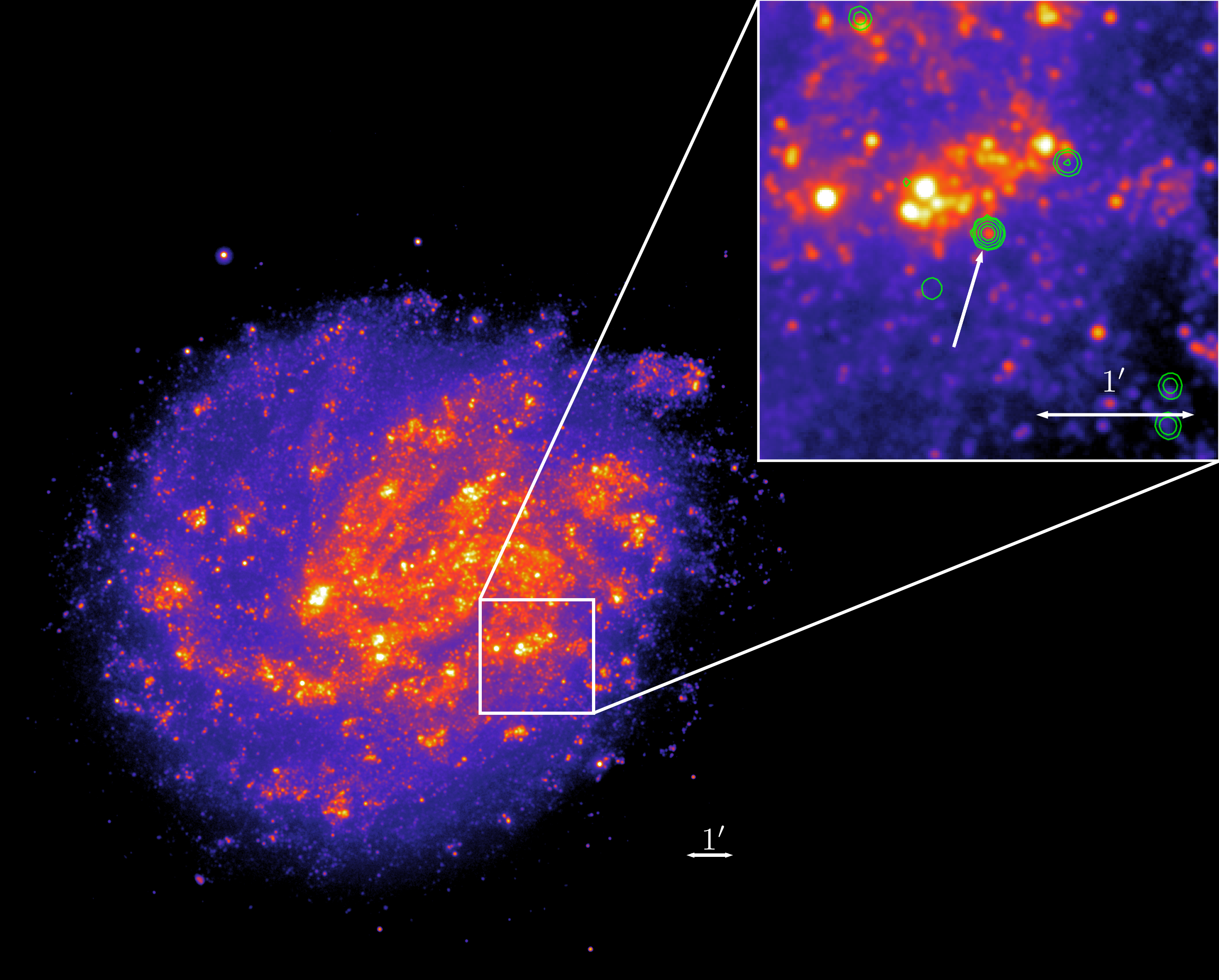}
    \caption{NGC\,300 from \swift/UVOT in the UVW2 band. The inset shows the location of \src, which is indicated by the white arrow. X-ray contours from the \cxo\ detection of November 2014 are plotted in green.\label{fig:ngc300}}
\end{figure*}

CXOU\,J005440.5--374320 (\src\ for brevity) was discovered in the CATS\,@\,BAR project with a large-amplitude flux modulation at a period of \mbox{$\sim$6\,h}. It is seen projected in the sky inside the inner region of the NGC\,300 (Fig.\,\ref{fig:ngc300}), a face-on spiral galaxy (Scd) in the Sculptor group at a Cepheid distance of 1.88\,Mpc \citep[][corresponding to a distance modulus of 26.4\,mag]{gieren05}.

NGC~300, due to its proximity and favourable inclination,  has been the subject of many studies about its stellar populations \citep[][and references therein]{bresolin09}, star formation history \citep{butler04}, Wolf--Rayet (WR) stars and OB associations \citep{schild03,pietrzynski01}, planetary nebulae \citep{pena12}, variable stars \citep{pietrzynski02,mennickent04}, dust content \citep{helou04,roussel05}, X-ray sources \citep{read01,carpano05,binder17,carpano18,urquhart19}, supernova remnants \citep{blair97,pannuti00,payne04,gross19}, and UV emission properties \citep{munoz07}. NGC~300 has been therefore \rev{observed with a wide variety of different telescopes} and a large amount of multiwavelength data is available.

In this work, we focus on the nature of \src. In Section \ref{observations}, we present the data sets and the observational properties of the source, highlighting its X-ray behaviour and its optical/UV counterpart. In Section \ref{discussion} we discuss possible scenarios which could originate \src\ emission and make our concluding remarks.

\begin{figure*}
\centering
\resizebox{\hsize}{!}{\includegraphics[angle=-90]{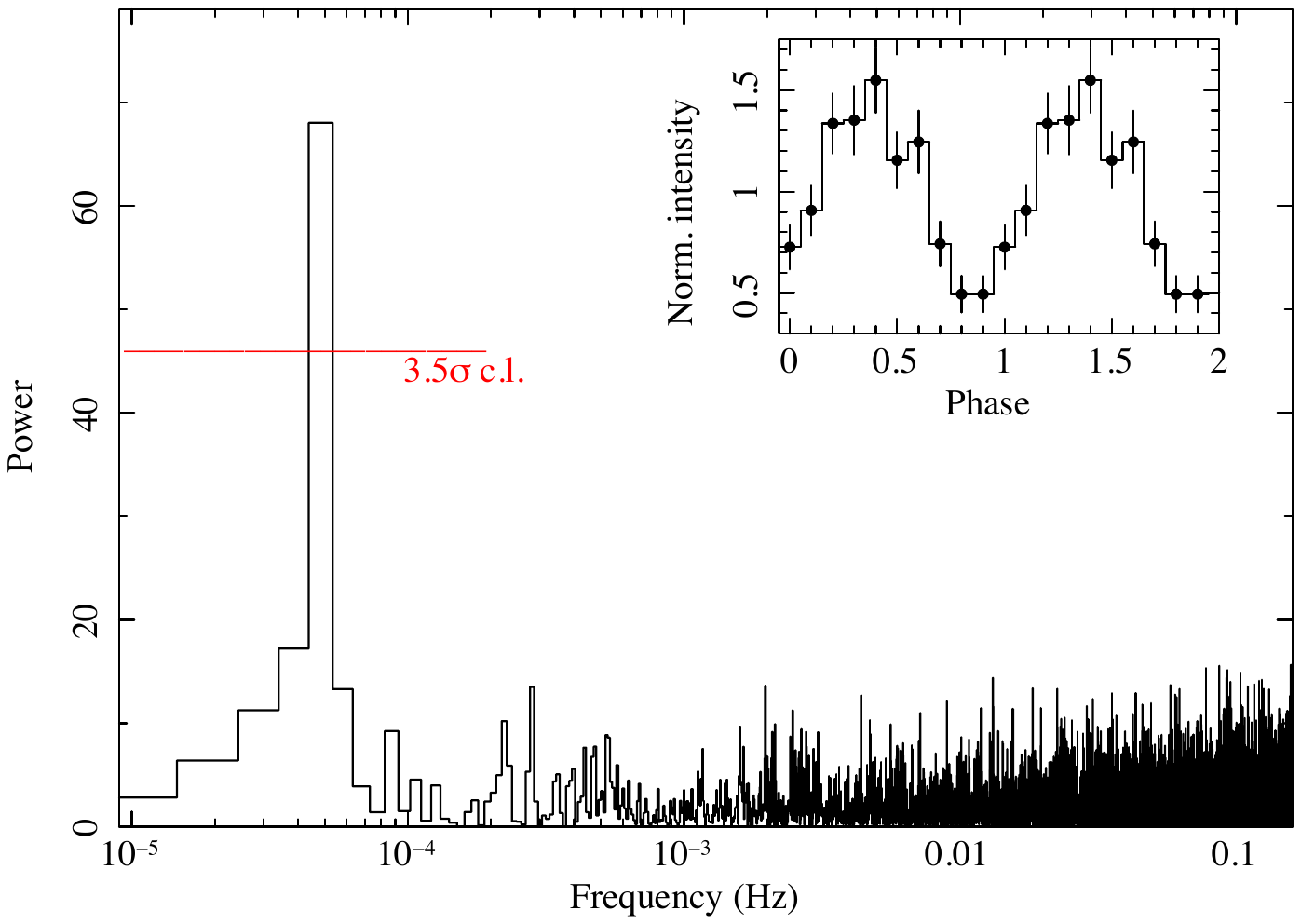}\hspace{-4cm}\includegraphics[angle=-90]{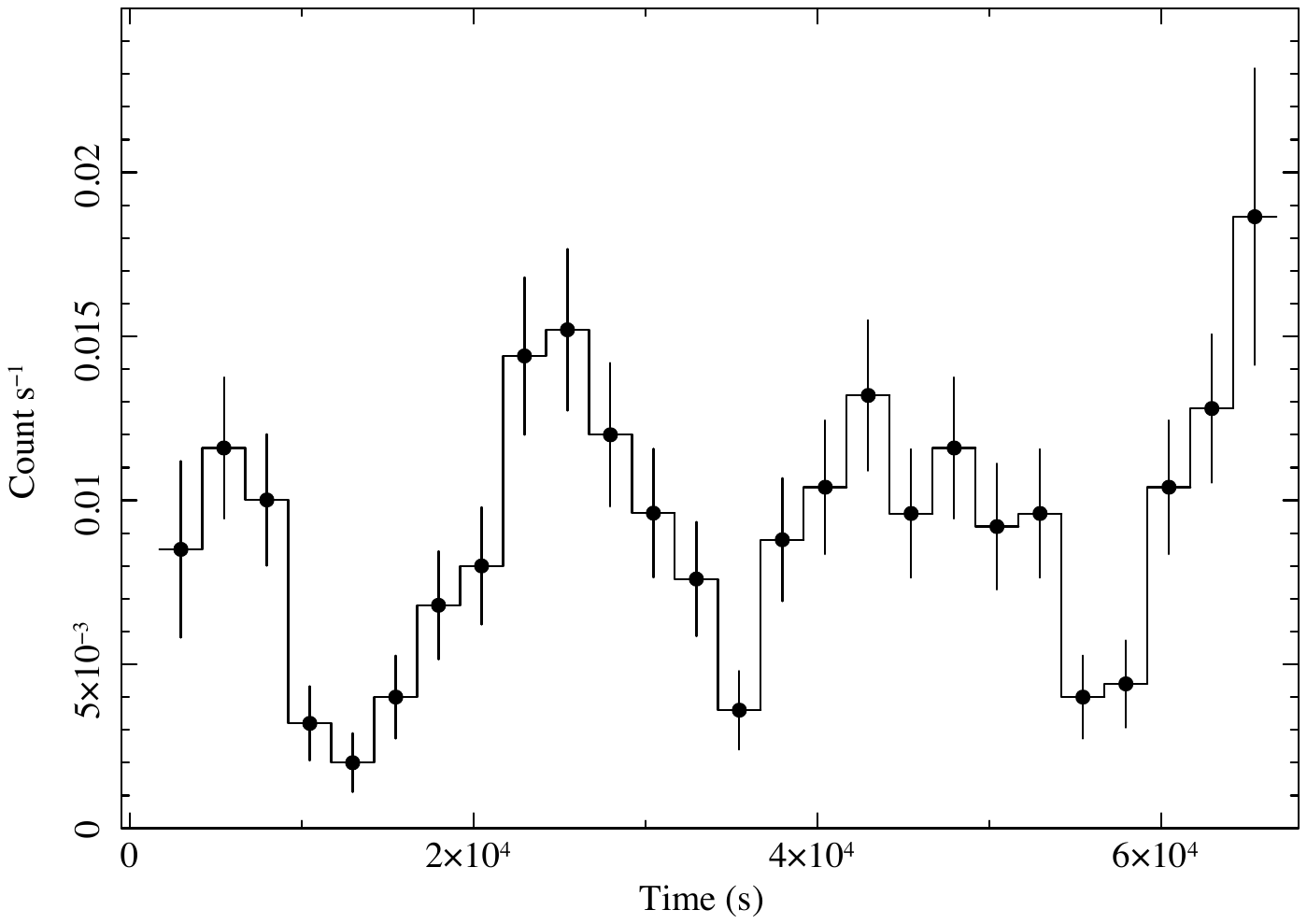}}
    \caption{Timing analysis of \src. \emph{Left panel:} Fourier power spectrum of the \cxo\ data of \src. The red line indicates the \rev{local} 3.5$\sigma$ detection threshold adopted in the CAT\,@\,BAR project. The prominent peak well above the threshold (\rev{5.9}$\sigma$) corresponds to the $\sim$6\,h modulation of \src. The inset shows the pulse profile obtained by folding the data at the best period of $5.88\pm0.12$\,h. \emph{Right panel:} Light curve (the background is negligible). The bin time is 2500\,s.
    \label{timing}}
\end{figure*}

\section{Observational properties}\label{observations}

\subsection{X-ray data sets}

\subsubsection{\cxo}
\src~fell into \cxo\ field of view four times, see Tab.\,\ref{tab:cxo}. Data were reprocessed and reduced with the Chandra Interactive Analysis of Observations software package (\texttt{CIAO}, v.4.12; \citealt{fruscione06}) and the \texttt{CALDB} 4.9.0 release of the calibration files. In the first two visits, the source was not detected and $3\sigma$ upper limits on its flux were obtained with the \texttt{CIAO} tool \texttt{srcflux}\revv{, following the default Bayesian approach}. In the two most recent observations, the source is located at 3.7 and 6 arcmins from the telescope aim point, respectively. The source counts were extracted \rev{from elliptical regions with semi-major
and semi-minor axes of 3.2 and 2.6 arcseconds for the November 2014
observation, and of 5 and 4 arcseconds for the April 2020 observation}, corresponding to a point-spread function (PSF) fraction of about 96\%. Backgrounds were estimated from \rev{source-free} annular regions, centred on the source position, of 10 and 20 arcseconds radii. 

\subsubsection{\swift/XRT}
Owing to the large monitoring campaign brought forth on NGC\,300 and its sources, the location of \src\ was in the \swift/XRT field of view on numerous occasions (170 times by the end of 2022) across a span longer than 15 years, but it was detected only in four visits: in August 2018, January 2019, April 2020, and one last time in September 2020, as reported in Tab.\,\ref{tab:cxo}.
Count rates, and $3\sigma$ upper limits for all the observations in which the source is not detected, were extracted using the \texttt{ximage} tool \texttt{sosta}\revv{, following the default Poissonian approach (a Bayesian approach was also tested and it provided slightly deeper upper limits, still however compatible with the Poissonian ones)}.
Merging together all \swift--XRT non-detections, the total exposure time amounts to $\approx570$ ks, but the source still cannot be detected. The $3\sigma$ upper limit on the stacked observation is $8.05\times10^{-5}$ cts\,s$^{-1}$, corresponding to a flux limit of $2.2\times10^{-15}$\,\flux\ (assuming the best fitting model described below).
 
\subsubsection{\xmm}
The position of \src\ was imaged with \xmm\ seven times, as reported in Tab.~\ref{tab:cxo}, but the source was never detected. For each observation, data were retrieved from the \xmm\ science archive and reduced following the standard procedure. 
The $3\sigma$ upper limits were obtained from the EPIC-pn cameras, with the sole exception of the fourth observation, in which the source falls on the EPIC-pn's border and hence data from the merged EPIC-MOS cameras were used. The merging of the MOS cameras was performed with the \texttt{sas} tool \texttt{merge}. The $3\sigma$ upper limits were obtained with the \texttt{eupper} tool, from a circular region of 15" radius centred on \src\ position, with background counts extracted from a nearby free-of-sources circular region of 30" radius on the same detector chip\revv{, following the default Bayesian approach}.

\subsection{Short-term variability}

The CATS\,@\,BAR pipeline detected the uncatalogued X-ray source \src\ in a $\approx$60\,ks exposure carried out in November 2014 (Table\,\ref{tab:cxo}) and singled it out as a new X-ray pulsator with a $\sim$6\,h flux modulation (Fig.\,\ref{timing}). Taking into account the 16\,377 independent  trials, the false alarm probability was $\sim$$7.6\times10^{-11}$, which corresponds to a 6.5$\sigma$ detection. \rev{After correcting for the (mild) red noise in the exposure following \citet{israel96}, we evaluated the significance of the signal at $5.9\sigma$.}
Due to the relatively poor statistics, the approach of using the light curve to estimate the significance of modulation is not recommended and results in an underestimation of the real statistical level. Nonetheless, as an additional test, we compared the fit of the lightcurve with a constant and a constant plus a sinusoidal component. This comparison gives a F-test probability of $5\sigma$ that the addition of the sinusoid is significant.

As a further test, we also simulated 10$^5$ Fourier power spectra with the same properties as that of \cxo\ (\revvvv{a Poissonian noise plus a $f^{-\alpha}$ additional component, where $\alpha=0.71\pm0.05$ is inferred by fitting the power spectra distribution continuum of the original \cxo\ dataset}) and verified that no significant peak was detected by the used detection algorithm (which takes into account for any additional non-Poissonian noise component) at frequencies shorter than 10$^{-3}$\,Hz, setting a probability threshold of \revvvv{10$^{-5}$ ($>$ 4.4$\sigma$)} at the 5.88h-period frequency. The above numbers are consistent with the 5$\sigma$ and the 5.9$\sigma$ level inferred by using two different independent approaches. Correspondingly, we conclude that the signal significance is larger than  $>4.4\sigma$ and likely lies in the 5--5.9$\sigma$ range.

From the fit of the sinusoidal function, we derived a period of \mbox{$P = 5.88\pm0.12$\,h}, which is entirely consistent with that reported in \citet[][]{israel16}. The pulsed fraction, estimated from the semi-amplitude of the sinusoidal function, was $52\pm4$\%. 
With a $\nu/\Delta\nu$ factor of $\sim$10, the signal is seemingly coherent (that is, strictly periodic), but this is something clearly difficult to assess when only a few cycles are sampled, and a quasi-periodic modulation cannot be excluded.

In the last \cxo\ visit, no significant pulsation was detected, although the 3$\sigma$ upper limit on the pulsed fraction is about 50\%. In the four \swift/XRT detections, the signal-to-noise ratio is too low to carry out a meaningful timing analysis.

\subsection{Long-term variability}

\begin{figure*}
	\includegraphics[width=\hsize]{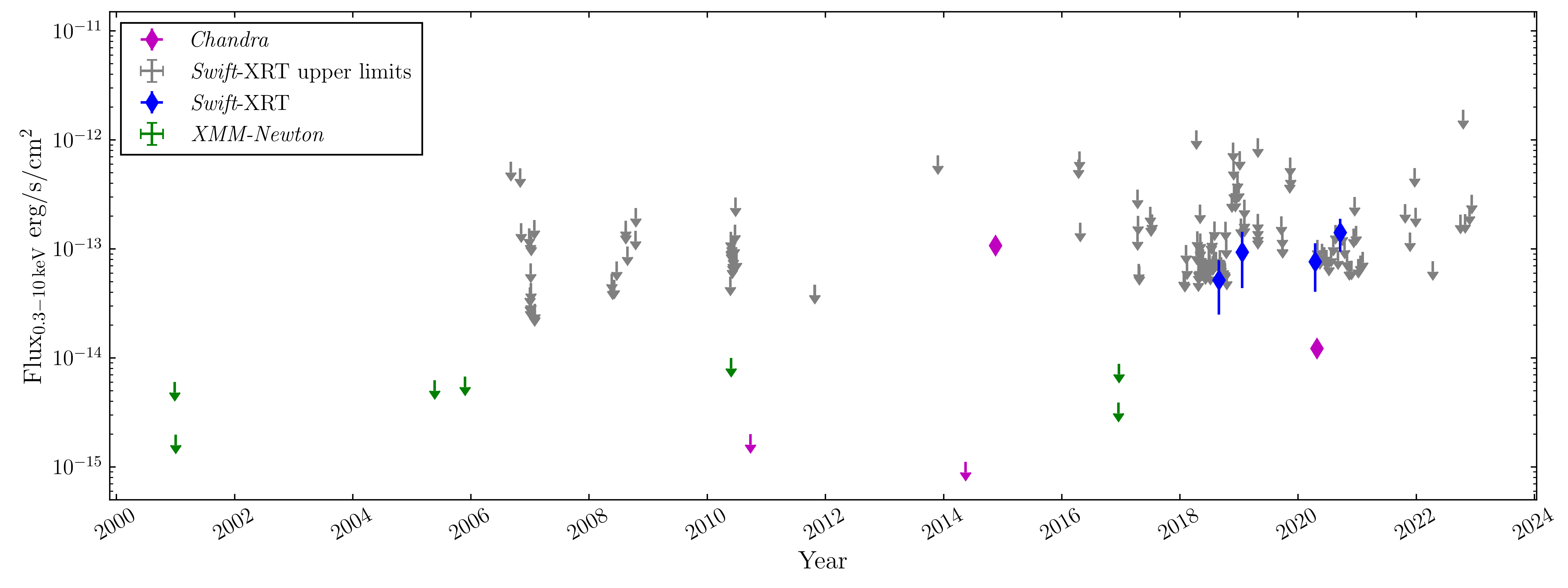}
    \caption{Long-term X-ray lightcurve of \src. Magenta and green data refer to \cxo\ and \xmm\ data, respectively; down-pointing arrows indicate $3\sigma$ upper limits while diamond markers indicate detections (with $1\sigma$ error bars). \swift/XRT detections are reported as blue diamonds, while the upper limits are indicated by down-pointing grey arrows. Count rates for the upper limits and for the \swift/XRT detections were converted to fluxes assuming the same spectral shape of the \cxo\ 2014 detection. Years are on the X-axis and flux in the 0.3--10\,keV band is on the Y-axis. \label{fig:longterm} }
\end{figure*}

Owing to the long-term monitoring of NGC\,300, we can trace \src\ X-ray variability across more than two decades (an early 1990s \emph{ROSAT} upper limit is unfortunately too shallow to be relevant in this regard).
\src\ was never detected before the \cxo\ pointing of November 2014 (Obs.ID 16029). The closest observation is the non-detection by \cxo, six months earlier, in May 2014 (Obs.ID 16028). The $3\sigma$ upper limit obtained in May 2014 implies a flux variation of at least a factor of 100.
After this, \src\ has not been pointed for $\sim$1.5 years and when NGC\,300 was observed again, it had disappeared: it is not detected in three short observations by \swift/XRT in April 2016 (although the upper limits for these observations are very shallow), nor in two longer observations by \xmm\ at the end of the same year.
\src\ reappeared in a \swift/XRT observation in late August 2018, unfortunately, the upper limits around this detection are too shallow to reconstruct the \rev{detailed} source behaviour.
The scene repeated on January 2019, April 2020 and September 2020. In April 2020 \src\ has been detected by both \swift/XRT (Obs.ID 00095672001) and \cxo\ (Obs.ID 22375) less than ten days apart.
After September 2020, \src\ has never been detected again, possibly also due to the visits on NGC\,300 growing sparser and shallower with time. Fig.\,\ref{fig:longterm} shows the long-term light curve of \src\ in the years from approximately 2000 to 2023, count-rates were converted into fluxes with the model described below.

\subsection{Spectral fitting}

Here we address the X-ray spectral fitting of \src. In particular, we focus on its first detection with \cxo\ in November 2014, which is the data set with the highest count statistics.

\begin{figure}
	\includegraphics[width=\hsize]{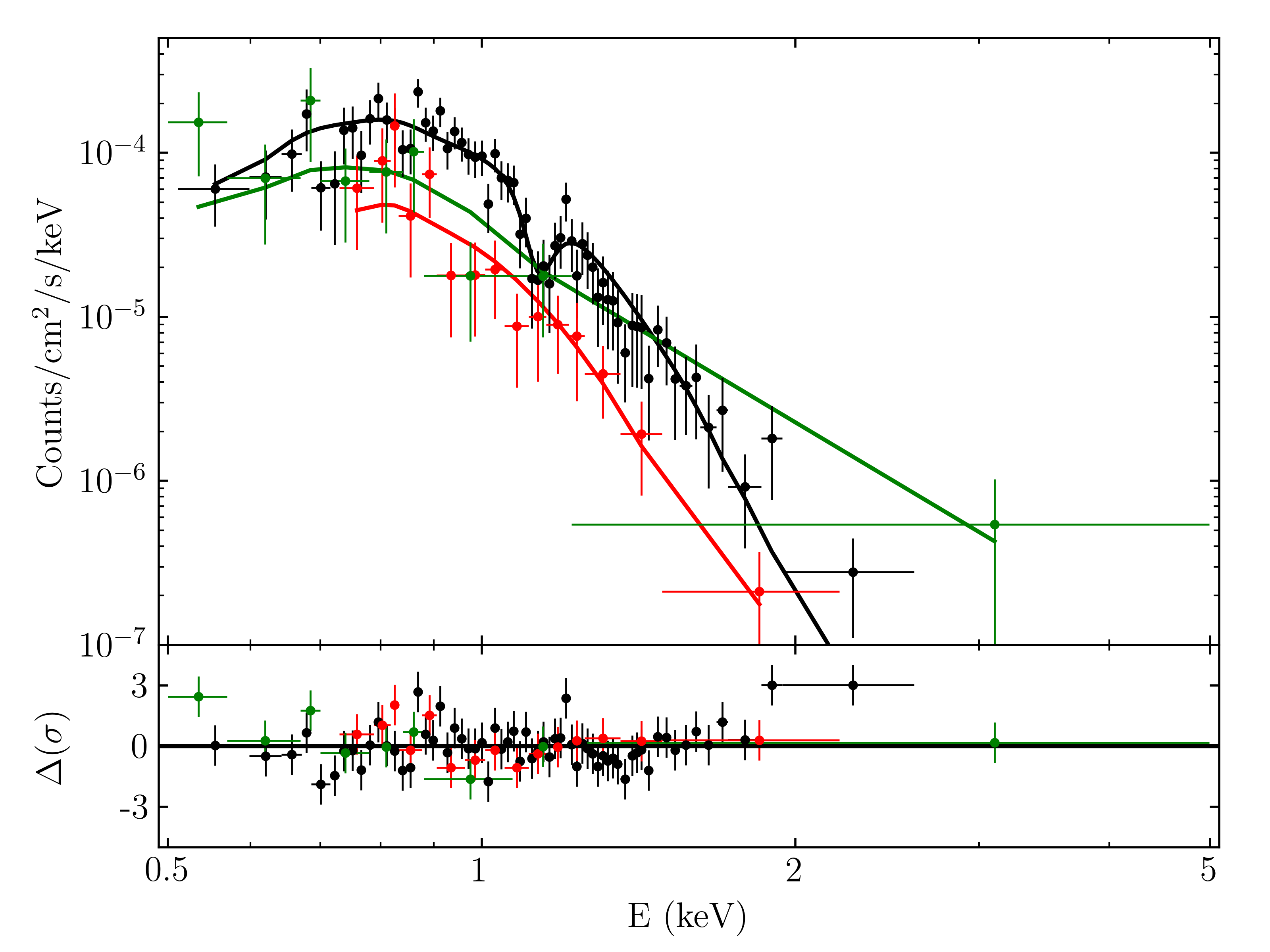}
    \caption{The upper panel shows the X-ray folded spectra and best-fitting model of \src. Residuals are shown in the lower panel. Data from \cxo\ November 2014, April 2020 and \swift/XRT merged detections in black, red and green dots, respectively. Solid lines indicate the best-fitting \textsc{slimd} model. \label{fig:spec}}
\end{figure}

The spectra, the spectral redistribution matrix and the ancillary response file were generated using the \texttt{CIAO} script \texttt{specextract}. The spectrum of \src\ was fed into the spectral fitting package \texttt{XSPEC} \citep{arnaud96} version 12.12.1. Spectral channels having energies below 0.5\,keV and above 7.0\,keV were ignored in the fit, owing to the very low signal-to-noise ratio from \src\ (essentially all the counts are between approximately 0.7 and 1.7\,keV). The spectra were grouped to have a minimum of 1 count per energy bin, and C-statistic \citep{cash79} was employed (for the spectra \rev{shown} in Fig.\,\ref{fig:spec}, further rebinning has been adopted for display purposes only).

We fit to the spectra a number of different simple models including (but not limited to) blackbody (\textsc{bbody}), power law, bremsstrahlung, Raymond--Smith plasma \citep{raymond77}, hot diffuse gas (\textsc{MEKAL}, \citealt{mewe85,mewe86,liedahl95}), collisionally-ionized gas (\textsc{APEC}, \citealt{smith01}), and accretion disc model with multi-blackbody components (\textsc{diskbb}\revv{, with standard colour-correction factor fixed at 1.7, \citealt{lorenzin09}}) all corrected for interstellar absorption (\textsc{TBabs}, with \nh\ fixed to the Galactic value $9.45\times10^{20}$ cm$^{-2}$; \citealt{hi4pi16}). The abundances used are those of \citet{wilms00}, with the photoelectric absorption cross-sections from \citet{verner96}.

While the power law, \textsc{MEKAL}, \textsc{APEC}, Raymond-Smith plasma and bremsstrahlung models provide unacceptable fits ($C/\nu$ between 2 and 3, \rev{$C$ being the value of the statistic and $\nu$ the number of degrees of freedom}); both the \textsc{bbody} and \textsc{diskbb} can reproduce well the spectrum, and the fits improved when an extra-layer of absorption and an absorption Gaussian line \revvv{at $1.13\pm0.01$ keV with width fixed to 0 and normalization $K=(-1.4\pm0.4)\times10^{-5}$ photons\,cm$^{-2}$\,s$^{-1}$, are added} \revvvv{(EW$=0.077\pm0.002$ keV)}. The significance of the extra layer of absorption and of the Gaussian line were tested with simulations. In both instances, the \texttt{Xspec} routine \texttt{simftest}\footnote{\texttt{simftest} was employed using default parameters and 10'000 iterations, details on the routine can be found at \url{https://heasarc.gsfc.nasa.gov/xanadu/xspec/manual/node126.html}} returned a $<0.01\%$ probability that the data are consistent with the model without the extra component. \revv{Furthermore, the statistic improvement of adding the line is $\Delta C=35.1$ for three degrees of freedom:} even accounting for the "look elsewhere" effect (e.g., \citealt{Lyons2008}), as the absorption feature does not line up with a specific atomic transition, by correcting the obtained p-value for the dimension of the energy space, \revvv{i.e. multiplying it for the ratio between the bandwidth and the average resolution: $(2~{\rm keV}$--$0.5~{\rm keV})/0.05~\rm{keV}\approx30$, the line significance results to be $>$$3.9\sigma$}. For both models, the best-fit parameters are reported in Tab.\,\ref{spec:params}.

\begin{table*}
\caption{X-ray spectral fitting parameters. \label{spec:params}}             
\centering                          
\begin{tabular}{lccccccc}        
\hline\hline    
model$^1$ & $N_\textup{H}^2$ & $kT$ & $K^3$ & $M_\textup{h}^4$ & $L_\textup{X}^5$ & $C/\nu^6$ & g.o.f.$^7$ \\   
\hline \\ 
\textsc{bbody} & $0.73_{-0.13}^{+0.16}$ & $0.109\pm0.006$ & $7.0_{-2.9}^{+6.6}\times10^{-5}$ & -- & $1.9_{-0.8}^{+1.3}$ & 80/80 & $26\%$\\
\textsc{diskbb} & $0.87_{-0.15}^{+0.14}$ & $0.117_{-0.003}^{+0.009}$ & $2.2_{-1.3}^{+3.8}\times10^{3}$ & 1.0 & $4.3_{-1.8}^{+1.6}$ & 79/80 & $27\%$\\
\textsc{slimd} & $0.9\pm0.1$ & -- & $0.11_{-0.01}^{+0.02}$ & $1.4_{-0.4}^{+1.0}$ & $3.6_{-0.9}^{+1.5}$ &  74/80 & 21\% \\
\hline                                   
\end{tabular}
\tablefoot{The errors correspond to a statistic variation $\Delta C=1$. $^1$The models are reported with their \texttt{Xspec} names. $^2$Intrisic absorption in units of $10^{22}$\,atoms\,cm$^{-2}$. $^3$Normalization of the model, corresponding to $K=L/D^2$ (with $L$ luminosity in units of $10^{39}$ erg/s and $D$ distance of the source in units of 10 kpc) for the \textsc{bbody} model, $K=(R^2/D^2)\cos\theta$ (with the internal radius $R$ in km and $\theta$ inclination of the disc, $\theta=0$ indicating face-on configuration) for the \textsc{diskbb} model, and $K=\dot m/\dot m_\textup{Edd}$ for the \textsc{slimd} model. $^4$ BH mass in units of $10^3\,M_\odot$. Derived from the model normalization for the  \textsc{diskbb} and as a fitted parameter for the \textsc{slimd} model, in both cases under the assumptions of Schwarzschild metric and $\theta=45^\circ$. $^5$Unabsorbed luminosity in units of $10^{39}$\,erg\,s$^{-1}$ in the 0.3--7 keV band. $^6$Value of C-statistic divided by the degrees of freedom. $^7$Goodness of the fit from Monte-Carlo simulations.} 
\end{table*}

\begin{figure}
	\includegraphics[width=\hsize]{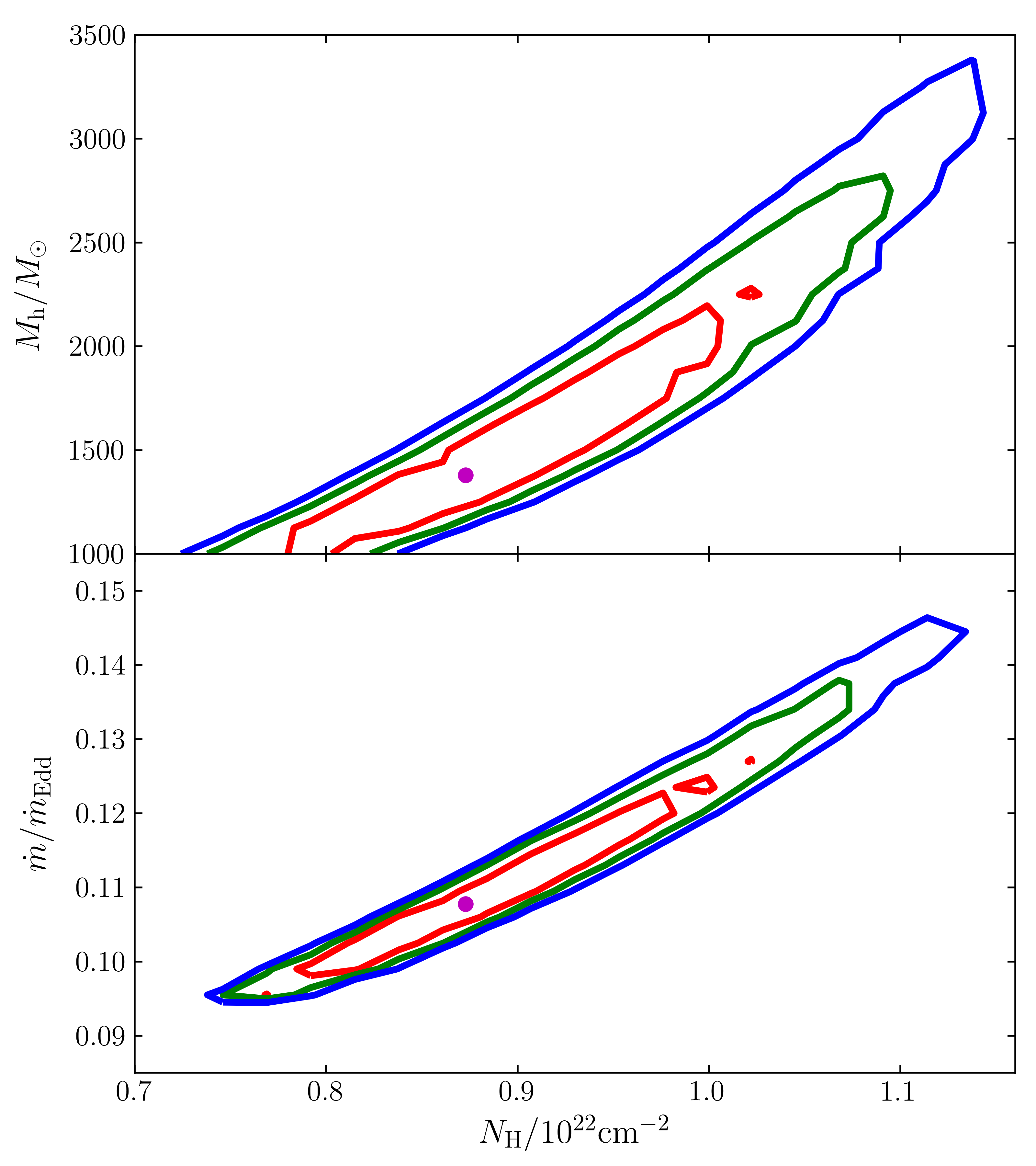}
    \caption{Contour plots in the BH mass--intrinsic absorption (upper panel) and mass accretion rate--intrinsic absorption spaces (lower panel)\rev{, obtained with the \textsc{slimd} model}. Red green and blue contours indicate the $\Delta C=1, 2, 3$ levels, while the magenta dots are the best-fitting parameters. \label{fig:cont}}
\end{figure}

Both the \textsc{bbody} and \textsc{diskbb} models provided similar temperatures, $kT\approx110-120$ eV, and similar amounts of intrinsic absorption, $N_\textup{H}\approx(0.7$--$0.8)\times10^{22}$\,atoms\,cm$^{-2}$. The \textsc{diskbb} model naturally suggests a black hole (BH) as the accretor. From the fit we get an emitting region of $\approx$9000 km, which corresponds, under the assumption of Schwarzschild metric and $45^\circ$ disc inclination, to a mass of about $1000\,M_\odot$. This model also naturally suggests the presence of an accretion disc. 
Since \src\ is a transient source, to investigate better this possibility, we tried a model designed to reproduce accretion discs around black holes in transient events, \textsc{slimd} \citep{wen22}. As in the previous cases, we added to the \textsc{slimd} model the two layers of absorption (one fixed to the Galactic value, as above, and the other free to vary) and an absorption Gaussian line at 1.13 keV, and kept the disc inclination and  spin fixed at $45^\circ$ and 0, respectively. 
We obtained good fit (Tab.\,\ref{spec:params}) for a BH with mass $1.4_{-0.4}^{+1.0}\times10^3\,M_\odot$ and moderate values of accretion rate ($\dot m=0.11_{-0.01}^{+0.02}\,\dot m_\textup{Edd}$, where $\dot m_\textup{Edd}$ indicates the maximum Eddington accretion rate) and intrinsic absorption $N_\textup{H}=0.9_{-0.1}^{+0.1}\times10^{22}$\,cm$^{-2}$, where errors correspond to a variation of the statistics $\Delta C=1$ (1$\sigma$). We note that both the BH mass and mass accretion rate heavily correlate with the intrinsic absorption, as shown by the contour plots in Fig.\,\ref{fig:cont}. The value of unabsorbed X-ray luminosity (between $0.3$ and $7$ keV), computed with this model and assuming \src\ is located in NGC\,300, amounts to $3.6\times10^{39}$ \lum, which is in the regime of the ultraluminous X-ray sources (ULXs, e.g. \citealt{kaaret17,pinto23}).

The value derived for the BH mass is affected not only by statistical uncertainties but also by systematic ones: we assumed fixed disc inclination and  spin, and of course, the estimate is model dependent. Furthermore, the \textsc{slimd} model does not support BH masses lighter than $1000\,M_\odot$, thus preventing us from fully exploring the parameter space, as highlighted by Fig.\,\ref{fig:cont}. In order to have an idea of these uncertainties, first of all we performed fits with different BH spin and disc inclination values: prograde rotation ($a_\bullet=0.99$) and edge-on ($\theta=90\degr$) provided heavier masses, up to some $10^4\,M_\odot$, although the maximally-rotating case ($a_\bullet=0.99$) provided an unacceptable fit ($C/\nu>3$). Counter-rotation ($a_\bullet=-0.99$) and face-on ($\theta=0\degr$) configuration provided, instead, lighter masses, but the \textsc{slimd} model would not allow us to explore the mass regime below $1000\,M_\odot$. To explore the lighter mass regime we employed the \textsc{slimbh} model \citep{sadowski11,straub11}. \rev{This is} designed to describe slim accretion disks around \rev{Eddington-limited} stellar-mass ($\lesssim$100\,$M_\odot$) BHs and has an upper limit for the BH mass of $1000\,M_\odot$, thus complementing the \textsc{slimd} model. The \textsc{slimbh} model with disc inclination fixed at $45^\circ$ provides an unacceptable fit ($C/\nu>3$) as the BH mass peaks at $1000\,M_\odot$. A face-on configuration provides an acceptable fit ($C/\nu=70/85$, 
 $10\%$ goodness of fit, \rev{i.e. the percentage of cases, out of 10\,000 spectra simulated based on the model, in which the test statistic was less than that of the data}) with a BH mass of $\approx 800\,M_\odot$. 
 Finally, we note that from the \textsc{diskbb} model, with the corrections by \citet{lorenzin09}, the mass can be estimated from 320 to 1000$\,M_\odot$ in a 90\% range for the inclination (0$\degr$--85$\degr$).
 All in all, although we can poorly constrain the BH mass, as it ranges from $\sim$300 to $10^4\,M_\odot$, the spectral results consistently indicate an IMBH.\\

Unfortunately, the spectra obtained in the other detections do not afford the possibility to perform meaningful spectroscopic analysis: the latest \cxo\  detection, in April 2020, is heavily affected by the loss of effective area of the ACIS camera below $0.7$ keV, while the four \swift/XRT detections, taken individually, have too low signal-to-noise ratio for a spectrum to be extracted. Nonetheless, we tentatively merged data from all four \swift/XRT detections with the \texttt{ftool} routine \texttt{extractor} and extracted a merged spectrum with \texttt{xselect} from a circular region of 20 arcsec radius centred on the source position. The background was extracted from a free-of-sources circular region of 1 arcmin radius at a 3 arcmin distance from \src\ location and arfs likewise merged. In such a way, we reached a total of 32 net counts with a source fraction of the total counts evaluated at 94.3 \%. We then fitted the best-fitting model we obtained above to the three spectra (the \cxo's from November 2014 and April 2020, and the \swift/XRT one from the combined data sets) simultaneously. We froze all parameters to the best-fit ones, except for the normalizations (for the \textsc{slimd} model, it corresponds to the mass accretion rate). We obtained an acceptable value of the statistic ($C=119$ with 145 degrees of freedom and goodness of 31\%) with no Gaussian lines in the latest \cxo\ observation and in the merged \swift/XRT spectra and mass accretion rates equal to $0.07$ and $0.09$ $\dot m_\textup{Edd}$ respectively. We can conclude that the emission of \src\ in the latest \cxo\ and in the \swift/XRT detections, is compatible with the one observed in the first \cxo\ detection, although spectral evolution or variations of the amount of intrinsic absorption cannot be assessed. Fig.\,\ref{fig:spec} show the three spectra and their residuals in black, red and green, respectively, as well as the best-fitting model.\\

\subsection{Optical/UV counterpart}

\subsubsection{UVOT photometry}\label{uvot}

As the location of \src\ was visited several times with \swift, we possess a large number of \swift/UVOT observations, \rev{which cover each of the 6 UVOT optical/UV bands}: V (5468 \AA), B (4392 \AA), U (3465 \AA), UVW1 (2600 \AA), UVM2 (2246 \AA), and UVW2 (1928 \AA). \rev{Based on visual inspection, we removed any observations in which a smoke-ring feature}, generated by the nearby foreground G8 star CD-38 301 \citep{cruz19}, was superimposed at the \src\ location.

\begin{figure}
	\includegraphics[width=\hsize]{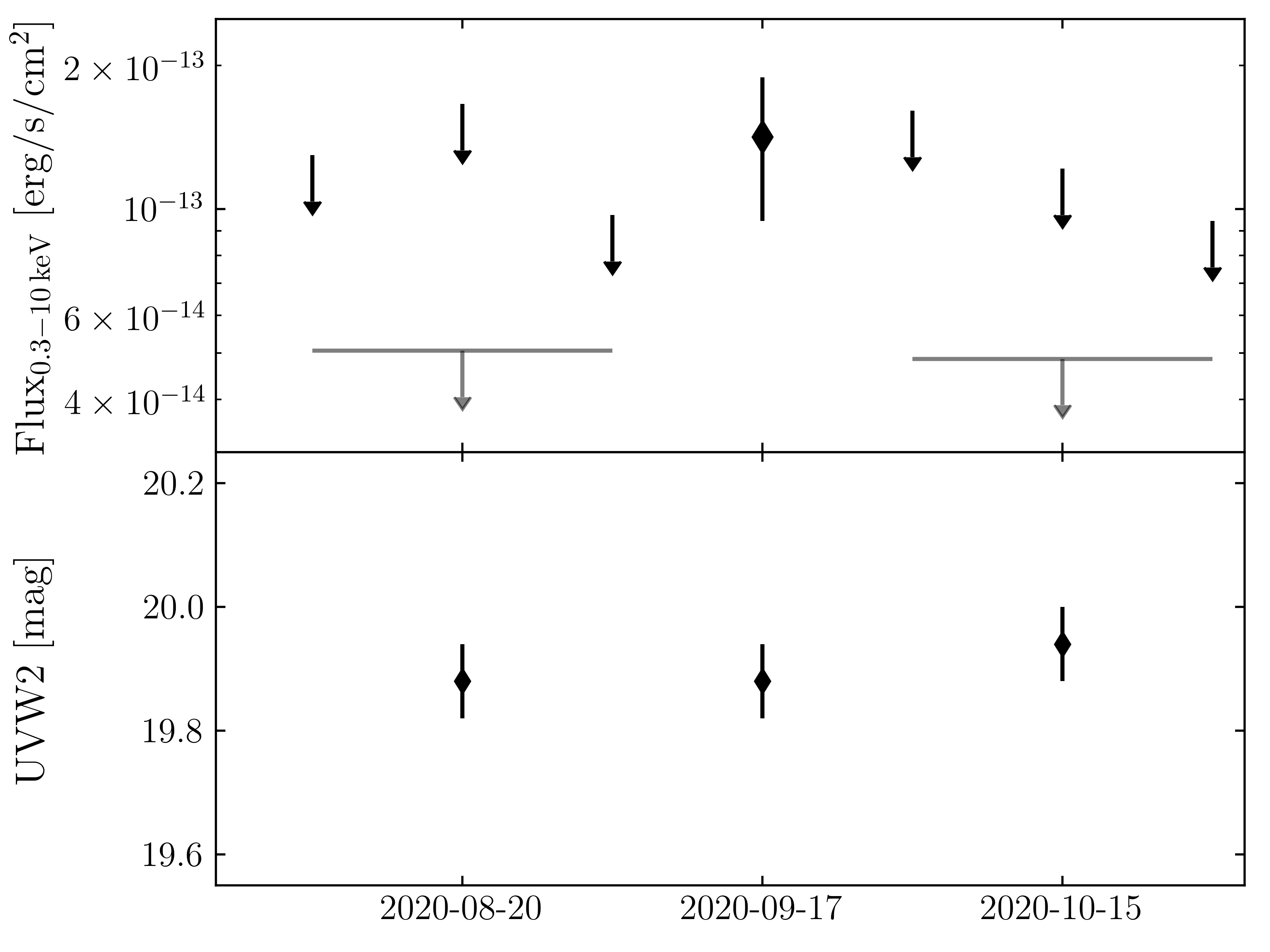}
    \caption{X-ray flux (upper panel) and UVW2 AB magnitude (lower panel) evolution during the flare detected in September 2020 \swift\ data. \revv{The X-ray upper limits in the single observations are reported as black arrows, the upper limits derived from the merged observations are reported as grey arrows.} \revvv{UVW2 magnitudes for each of the three epochs were derived by merging the single orbit exposures.} \label{fig:mw_var}}
\end{figure}

We found no trace of \rev{significant} variability \rev{(within $1\sigma$, roughly corresponding to half a magnitude)} in any of the six bands; in particular, we found none in coincidence with the X-ray flaring activity, as shown in Fig.\,\ref{fig:mw_var}, which highlights the difference between the UVW2 AB magnitude and the 0.3--10\,keV flux at the epoch of the latest flare. 
As no variability was detected, all observations were stacked using the \texttt{ftool} routine \texttt{uvotimsum} and fluxes were extracted using \texttt{uvotsource} (with flag APERCORR=curveofgrowth) from a circular region centred on the source position and with a radius of 5 arcsec. Backgrounds were estimated from a circular region of 10 arcsec radius, free of sources, at approximately 15 arcsec from the source location. The AB magnitudes from the merged \swift/UVOT observations are reported in Tab.\,\ref{tab:uvot}, \revvv{as well as the averaged magnitudes over the single observations}.

In order to correct the \swift/UVOT spectral energy distribution (SED) for dust extinction we can adopt different strategies. The minimum possible value of the reddening is the Galactic line-of-sight value $E(B-V)=0.0108$ \citep{schlafly11}: in this case, the optical/UV SED would be consistent with a blackbody profile of $\approx$15,000\,K and a luminosity of $7.5\times10^{4}\,L_\odot$, compatible with a \rev{blue supergiant star ($M_{\rm V} \approx -6.5$ mag).} A slightly higher reddening was adopted by \citet{gieren04}, who assumed a foreground Galactic $E(B-V)=0.025$ mag \citep{burstein84} plus a reddening $E(B$--$V)=0.05$ mag through the halo of NGC\,300. In this case, the best-fitting blackbody temperature rises to about 17,000\,K and the luminosity to $10^5~L_\odot$, still compatible with a single \rev{blue supergiant star ($M_{\rm V} \approx -6.7$ mag). 

However, the complete absence of individual stellar lines in the observed optical spectra (see Sec. \ref{salt}) makes it unlikely that the observed SED stems from a barely reddened, single blue supergiant or a (very) late-type WR star with a comparable temperature \citep[e.g.][]{sander14}.} 

The other extreme in terms of possible extinction would be to use the $N_{\rm H}$ inferred from the X-ray spectral modelling. Converted to an optical extinction $A_\textup{V}$ via the empirical relation in \citet{foight16}, this would imply $A_{\rm V} \approx 3.1$ mag, and hence an absolute brightness $M_{\rm V} \approx -9.6$ mag. Such a source would be too luminous to be a single star \rev{and would further be unphysically blue ($B-V\approx -1.2$ mag) for {\it any} star, cluster, or blackbody-like spectrum. Thus, we can rule out such a high extinction for the optical counterpart. Assuming a physically meaningful limit of B$-$V $\gtrsim -0.5$ mag and $U-B\gtrsim-1.5$ mag, we can place an upper limit on the total extinction of $A_{\rm V}\lesssim1$ mag. In this case, the photometry is compatible with a young stellar cluster (YSC) with an age of $\lesssim4$ Myr and $\approx10^3 M_{\odot}$, assuming instantaneous star formation and a Large Magellanic Cloud (LMC)-like metallicity (similar to the one of NGC~300\revv{:} \citealt{gazak15,cipriano16}). This would correspond to a small stellar population dominated by 3 or 4 main-sequence O stars and about $12-15$ B stars (see Sec. \ref{salt}).}

\begin{table*}
\caption{UV/optical magnitudes from \swift/UVOT and \xmm/OM. \label{tab:uvot}}             
\centering                          
\begin{tabular}{lccccccc}        
\hline\hline    
\multirow{2}{*}{Instrument} & \multirow{2}{*}{Epoch} &\multicolumn{6}{c}{AB Magnitude} \\
 & & V & B & U & UVW1 & UVM2 & UVW2 \\   
\hline 
\swift/UVOT (merged) & --& $19.84\pm0.09$ & $19.66\pm0.05$ & $19.75\pm0.03$ & $19.93\pm0.04$ & $20.06\pm0.05$ & $20.16\pm0.05$ \\
\swift/UVOT (averaged) & -- & $19.82\pm0.11$ & $19.56\pm0.09$ & $19.77\pm0.07$ & $19.79\pm0.07$ & $19.85\pm0.06$ & $19.92\pm0.05$\\
\hline
\multirow{6}{*}{\xmm/OM}& 2000-12-26 & -- & $20.2\pm0.1$ & -- & -- & -- & -- \\
& 2001-01-01 & -- & $20.1\pm0.1$ & $20.3\pm0.1$ & $19.5\pm0.1$ & -- & -- \\
& 2005-05-22 & -- & -- & $20.0\pm0.1$ & $19.9\pm0.1$ & $20.2\pm0.1$ & -- \\
& 2005-11-25 & -- & -- & $20.0\pm0.1$ & $19.62\pm0.05$ & $20.1\pm0.1$ & -- \\
& 2016-12-17 & -- & -- & -- & -- & $20.3\pm0.1$ & -- \\
& 2016-12-19 & -- & -- & -- & -- & $20.1\pm0.1$ & -- \\
\hline                            
\end{tabular}
\tablefoot{\swift/UVOT AB magnitudes from the merged observations \revvv{and averaged over the single ones}. \xmm/OM magnitudes are from the \xmm/OM catalogue \citep{page12}. U and UVW1 magnitudes from the 2016 \xmm/OM observations have not been reported as they are affected by the "smoke ring" feature. Quantities are reported with $1\sigma$ errors.} 
\end{table*}

\subsubsection{Archival data}

\src\ is also present in the \xmm/OM catalogue of serendipitous sources \citep{page12}. Our source location was visited six times as reported in Tab.\,\ref{tab:uvot}. \revV{\xmm/OM data are compatible with the \swift/UVOT ones when one considers that, at the magnitudes of our source, discrepancies as large as 1 magnitude are not uncommon between \swift/UVOT and \xmm/OM (see \citealt{yershov14} and the \xmm/OM calibration report\footnote{\url{https://www.cosmos.esa.int/documents/332006/2169043/calibration_om.pdf}}).}
\src\ is also present in the \gaia\ catalogue \citep{gaia23k}, with a parallax $p=0.53\pm0.78$\,mas, corresponding to a distance of $1.2_{-0.4}^{+0.5}$ kpc \citep{lindegren21,bay21}. However, at a magnitude $G = 20.49$ ($G_\textup{BP}=20.61$ and $G_\textup{RP}=20.37$), the source is rather faint for \gaia\ and the low signal-to-noise ratio parallax is probably unreliable.

\subsubsection{SALT spectrum}\label{salt}

Two spectra were acquired on Dec. 2, 2019 and Dec. 15, 2019 (Progr.ID: 2018-2-LSP-001) with the {\em Southern African Large Telescope} (SALT) \citep{buckley06, odonoghue06},  equipped with the  Robert Stobie Spectrograph (RSS) \citep{burgh03,kobulnicky03} in the long-slit ($8'\times1\farcs5$) spectroscopy mode. The PG0300 and PG0900 gratings were used for the observations with exposure times of 1800~s and 2000~s, respectively. A grating tilt of $5\fdg4$ was used for the PG0300 observation,  which covered the wavelength range 3700--7500 \AA\ and provided a resolving power of 250--600. For the PG0900 observation, a tilt of $15\fdg125$ (4200--7250 \AA) was used, with a resultant resolving power of 800--1200. The position angle (PA) was 0 degrees from North and the average seeing during the observations was $1\farcs5$.

Standard spectral reduction (bias, flat-field, sky subtraction and cosmic-ray removal) was performed using PySALT pipeline \citep{crawford10}. The wavelength calibration was performed with the Ar (PG0300) and Xe (PG0900) lamps and the 1D spectrum was extracted using various tasks in IRAF. Since no spectrophotometric standard star was observed, the spectra are not calibrated in flux. 

\begin{figure}
    \includegraphics[height=\hsize, angle=-90]{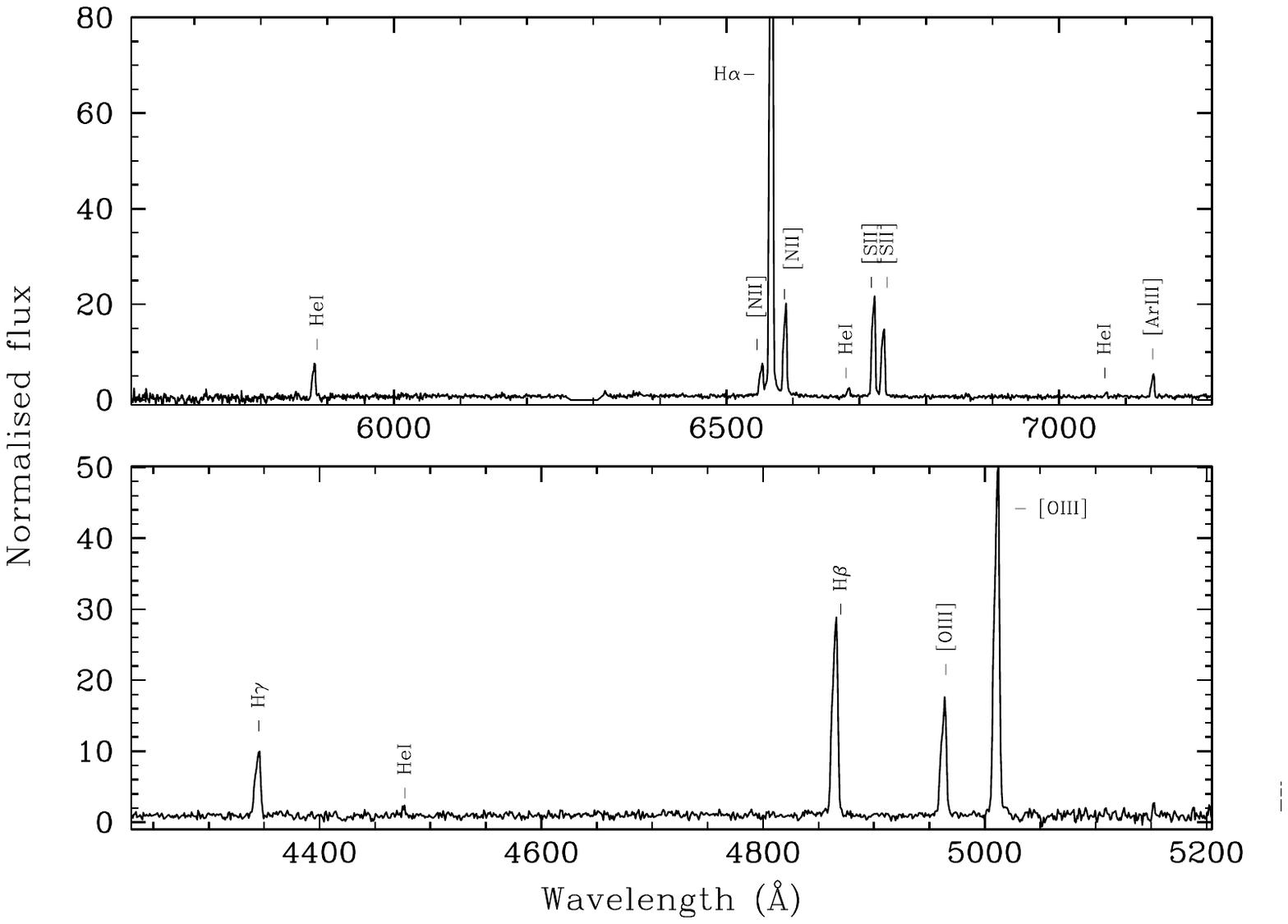}
    \caption{The SALT spectrum (PG0900) of \src. \label{fig:salt}}
\end{figure}

The two spectra show similar features and display strong emission lines of Balmer series and neutral He, as well as forbidden lines of [O\textsc{ii}], [O\textsc{iii}], [Ne\textsc{iii}], [N\textsc{ii}] and [S\textsc{ii}], with a blueward asymmetry, probably due to composite components in the emitting region, which however cannot be resolved with this resolution (see Fig.\,\ref{fig:salt}; the lower-resolution spectrum is shown in Fig.\,\ref{fig:salt_appendix} and the parameters of the main lines, which were derived from Gaussian fits, are given in Tab.\,\ref{tab:salt}).


The lack of flux calibration does not allow us to measure emission line fluxes ratios. \rev{However, we can use the broadband SED from \swift/UVOT and the unabsorbed blackbody fit described in Sec. \ref{uvot}, to estimate an approximate continuum flux level (assuming that it did not change substantially between the \swift\ and SALT observations), and then convert the measured line equivalent width (EW) to line fluxes. In particular, we used the P0900 grating spectrum from 2019 December 15, which provides the more reliable measurement of the line profiles because of its higher resolution.}

\rev{We find an observed flux ratio $F({\rm H}_\alpha)/F({\rm H}_\beta) \approx 2.9$ and  $F({\rm H}_\gamma)/F({\rm H}_\beta) \approx 0.47$, perfectly consistent with photo-ionized gas in HII regions (Case B recombination) without any reddening corrections apart from the line-of-sight Galactic component. Our simulations with the {\sc starburst99} code \citep{leitherer99,leitherer14} show that the EWs of H$_{\alpha}$ and H$_{\beta}$ are consistent with those expected from photo-ionized gas illuminated by a YSC with the same continuum flux we observed. Both of those findings suggest that the emission lines are from an extended ionized nebula rather than from the direct stellar counterpart of the X-ray source. Moreover, the ionized gas sees the same continuum flux as we observe. If there is additional intrinsic extinction, it must be located between the X-ray source and the surrounding nebula, rather than between the source of the line emission and the line of sight.}

\rev{We note that an O-type ionizing star embedded in a H\textsc{ii} region is not supported by the lack of absorption features of He\textsc{ii} and He\textsc{i} \citep{mcleod15,mcleod20} while only He\textsc{i} in emission is observed.}

\rev{If the 6-hr periodicity identified in the X-ray lightcurve corresponds to the orbital period of a binary system, a compact, WR-like donor is needed: assuming a total mass of the system of $\approx30\,M_\odot$ and a mass ratio of $\approx3$, the binary separation amounts to $\approx5\,R_\odot$ and the upper limit on the radius of the companion to $\approx2.5\,R_\odot$. However, our SALT spectra do not show any explicit WR features, such as broad He\textsc{ii} 4686 \AA\ emission (e.g. \citealt{dodorico83}). Also, the lack of a number of diagnostic lines \citep{schild92} does not support a WC \revv{or a WN: C\textsc{iii}, C\textsc{ii}, or C\textsc{iv} for the former and N\textsc{iii} 4100 \AA\ and 4640 \AA\ for the latter, are not detected}. While this could in principle be explained by a compact (``stripped'') star that does not show WR-type spectral features \citep[e.g.][]{goetberg18}, such stars are a huge source of He\textsc{ii} ionizing flux \citep[e.g.][]{sander20,sander23}, which should result in nebular He\textsc{ii} emission, which is absent as well. We thus ran {\sc starburst99} simulations to test whether stellar line emission features from a WR star could be hidden in a YSC with the observed continuum brightness and the line emission from its surrounding H\textsc{ii} region. Assuming the same uniform extinction $A_{\rm V} \approx 1$ mag for both the YSC and a WR star inside it, we verified that all features of a typical compact WR would be sufficiently diluted at least for some classes of WRs. For the WR, we used LMC models from \citet{hainich14} calculated with the PoWR \citep{graefener02,hamann03,sander15} model atmosphere code and diluted to the distance of NGC~300. From our calculations, we can thus conclude that the lack of observed WR emission lines in the SALT spectra does not rule out the scenario of a WR donor for the X-ray source. Moreover, an additional viable scenario is that the extinction around the X-ray binary system (including a WR donor with a dense wind) is higher than the extinction of the surrounding group of OB stars, which would further dilute or completely hide strong WR emission lines.}


\section{Discussion}\label{discussion}

\begin{figure}
	\includegraphics[width=\hsize]{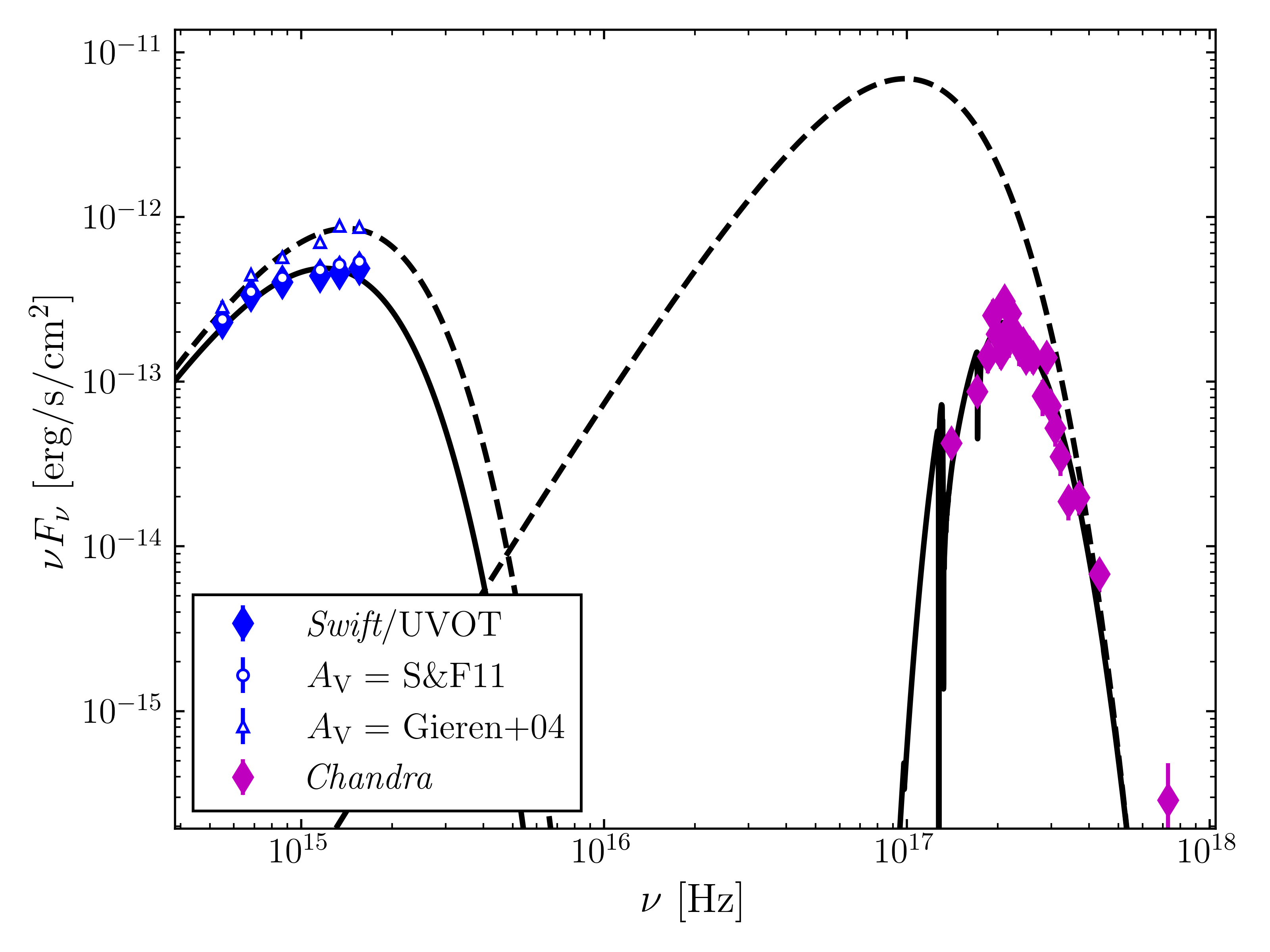}
    \caption{\revv{Full SED of \src. Magenta and blue diamonds show the \cxo\ and \swift/UVOT data respectively. The black solid lines show the two independently best-fitting models for the two components with no absorption correction, and the dashed black lines show the de-absorbed models. For \swift/UVOT, the full markers indicate the observed flux, while the empty blue markers the de-absorbed fluxes, corrected following the different values discussed in the text (the full markers lie under the empty ones). For \cxo, de-absorbed model was obtained by putting to 0 the amount of $N_\textup{H}$.} \label{fig:SED}}
\end{figure}

\src\ is a puzzling source. It was discovered as a bright X-ray pulsator with a 6\,h modulation, it shows a peculiar supersoft X-ray spectrum and displayed variability also on long-term. The available spectroscopy and photometry for the optical/UV counterpart show no hints of variability, but reveal only an HII region. While the photometry as such could match an individual B-star the strong absorption in the X-ray regime indicates that the photometry reflects an unresolved population of stars. Thus, any B-star, O-star, WR-star or even a whole YSC are viable options. \revv{In any case, the optical/UV and X-ray cannot be fitted simultaneously by a single black body or disk model, \revvv{nor it is possible to intercept the optical/UV data extrapolating the model best-fitting the X-ray and vice versa (even accounting for the parameters' uncertainties)}, as clearly demonstrated by Fig. \ref{fig:SED}, which shows the full SED and the absorbed and de-absorbed models. This suggests that the two emissions, optical/UV and X-ray, have different origins.}
Here we discuss different scenarios which could explain its peculiar multiwavelength appearance.

Projected in the sky, \src\ is seen well inside the B-$R_{25}$ radius\footnote{The 25$^{\rm{th}}$-magnitude isophote in the blue B band \citep{devaucouleurs91}.} of NGC\,300, at only 3.4\,arcmin (about 1.9\,kpc) from the center of the galaxy. Also, the Galactic latitude is $b\simeq-79\fdg4$, making a Galactic foreground object highly unlikely.
The only piece of information, possibly suggesting a foreground Galactic object, is the \gaia\ parallax, which is, however, probably unreliable given its low S/N.
In the hypothesis of a Galactic object, in view of the 6\,h modulation, the pulse profile and the flux, a cataclysmic variable (CV) is the only plausible candidate (although the X-ray spectrum discourages this interpretation and CVs are very unlikely to occur at the latitude of \src; e.g. \citealt{drake14}). \rev{If the X-ray period traces the binary motion, for a CV in a 6-h orbit, a donor of spectral type of K7V \citep{knigge06} is expected}. If we assume for it $M_\textup{V}=7.7$, $B-V=1.3$, and $V-R=1.15$ \citep[e.g.][]{johnson66,gray09}, the possibility appears unlikely, since a late type K7V would be too bright and red with respect to our source optical counterpart.
In the case of weakly-magnetic or non-magnetic system, the relationship between luminosity and orbital period by \citet{warner87} indicates absolute magnitudes of $V\approx4$ in a high state or $\approx$8 while in quiescence. More in general, the absolute magnitudes of such systems are always between 8 and 2 \citep{ramsay17} and are therefore incompatible with the optical counterpart of \src.

The lack of correlation between the X-ray flares and the optical/UV emission and the overall lack of significant variability at optical/UV wavelength across years does not support a nova at later stages \citep{williams94,williams16} nor a super-soft source (SSS; \citealt{vandenheuvel92}).
A SSS scenario on the other hand is also discouraged by the fact that the observed X-ray and optical luminosity would require a massive WD, of about $1.2\,M_\odot$ \citep{starrfield04} and this would place \src\ at $\approx50$--200\,kpc, which would result in an isolated intergalactic SSS, a highly unluckily occurrence.

\rev{The most natural way to explain the multi-wavelength behaviour of \src\ is probably an ultraluminous supersoft X-ray source (ULSs) scenario. ULSs can be considered as high-inclination, high-accretion rate compact objects. Here, the hard X-ray photons coming from the inner regions are significantly blocked by a strong optically-thick wind, which manifests itself in the form of Doppler-shifted absorption lines similar to the feature shown at 1.13 keV in Fig. \ref{fig:spec}\, \citep{pinto23,urq16}. Moreover, this feature seems to be a hallmark strongly associated with the ULS population \citep{urq16}. 

A well-known example is NGC\,247 ULS or X--1 for which high-resolution spectra resolved the 1 keV feature into a forest of lines produced by a powerful $0.17c$ wind \citep{pinto21}; this source also exhibits modulations on timescales of a few ks \citep{alston21}. Here, the spectral-timing analysis would rather support the case for a heavy NS or a small stellar-mass BH accreting well beyond the Eddington limit \citep{dai21}.}

\src\ also shares some spectral characteristics of M\,101\,ULX-1, a well-studied ULX and supersoft source consisting in a stellar-mass BH and a WR star \citep{liu13}. Indeed, the only object compatible with the 6\,h modulation, if it reflects the binary orbit, is a WR (see \citealt{eism13,eimm15,qiu19} for the discussion of similar systems and candidates). As discussed in Sect.\,\ref{salt} a WR, isolated or belonging to a YSC, is not excluded by the optical information. While for the compact object, it is unsafe to rule out a neutron star only on the basis of the observed super-Eddington luminosity (e.g. \rev{\citealt{bachetti14,israel17}}), a stellar-mass BH is the most likely compact object in WR binaries for evolutionary reasons \citep{vandenheuvel17}.
Both sources exhibit a highly variable (by a factor of $\approx$300) X-ray emission with a similar spectrum (at least in some states of M\,101\,ULX-1). In particular, similar temperature, size, and a dip at 1.1 keV \citep{soria16,urq16}. 

\rev{Finally, it is worth mentioning that the NGC~300 galaxy hosts another bright source, a.k.a. NGC 300\,X--1, which hosts a WR star feeding a massive black hole candidate (with a dynamically-inferred mass of $20\pm4\,M_{\odot}$, although a lighter compact object cannot be \revv{completely} ruled out; \citealt{Laycock2015}) which sometimes crosses the $10^{39}$\,\lum\ threshold \citep{Crowther10,Earnshaw17}.}

\revv{The fact that the absorption inferred from the X-ray spectrum is local to the X-ray source and does not involve the optical counterpart (as demonstrated in Sect. 2.5.1) might support an outflowing wind in this scenario. However, the } 
main problem with a ULX/ULS scenario is probably the transient nature of the source as, with a compact object in such a tight orbit with a strongly-winded WR, one would always expect significant accretion. \rev{A possible solution to this would be interpreting the variability in terms of obscuration, which would, however, likely result in spectral changes. The available   data, unfortunately, are inconclusive about this matter.} \revv{Finally we note that this source is not present in any ULX/ULS catalogues \revvv{(e.g. \citealt{walton22})} because, although its unabsorbed luminosity exceeds the $10^{39}$ \lum\ threshold, its observed, absorbed luminosity does not.}

A radically different scenario, which is capable of explaining some---although not all---features of \src, is the partial tidal disruption of a star by an intermediate-mass BH (IMBH). 
Tidal disruption events (TDEs) occur when a star wanders too close to a supermassive BH (SMBH)---usually---and its binding self-gravity gets overcome by the BH's tidal forces, tearing the star apart. Part of the stellar debris gets subsequently circularized and accreted, emitting a bright electromagnetic (EM) signal, and part gets ejected \citep{ree88, phi89}. TDEs, although being bright across all EM spectrum, from radio to $\gamma$-rays (e.g Swift J1644, \citealt{bur11}), in the X-ray band usually appear as luminous super-soft transients. 
Indeed, the X-ray spectrum suggests that \src\ emission is linked to accretion processes and the \textsc{slimd} model (specifically developed to fit slim disk emission from the tidal disruption of stellar objects by SMBHs) provides an estimate for the mass of \src\  
($\approx $1400\,$M_\odot$) placing it in the IMBH class.
The TDE hypothesis could also explain the observed long-term X-ray variability. As invoked for XMMU\,J122939.7+075333 \citep{tie22}, an object sharing with \src\ many characteristics (long-term variability, very soft X-ray emission, forbidden lines in the optical counterpart), a star on a highly eccentric orbit would get stripped of its material at every pericentre passage, thus producing recurrent flares. 

\begin{figure}
	\includegraphics[width=\hsize]{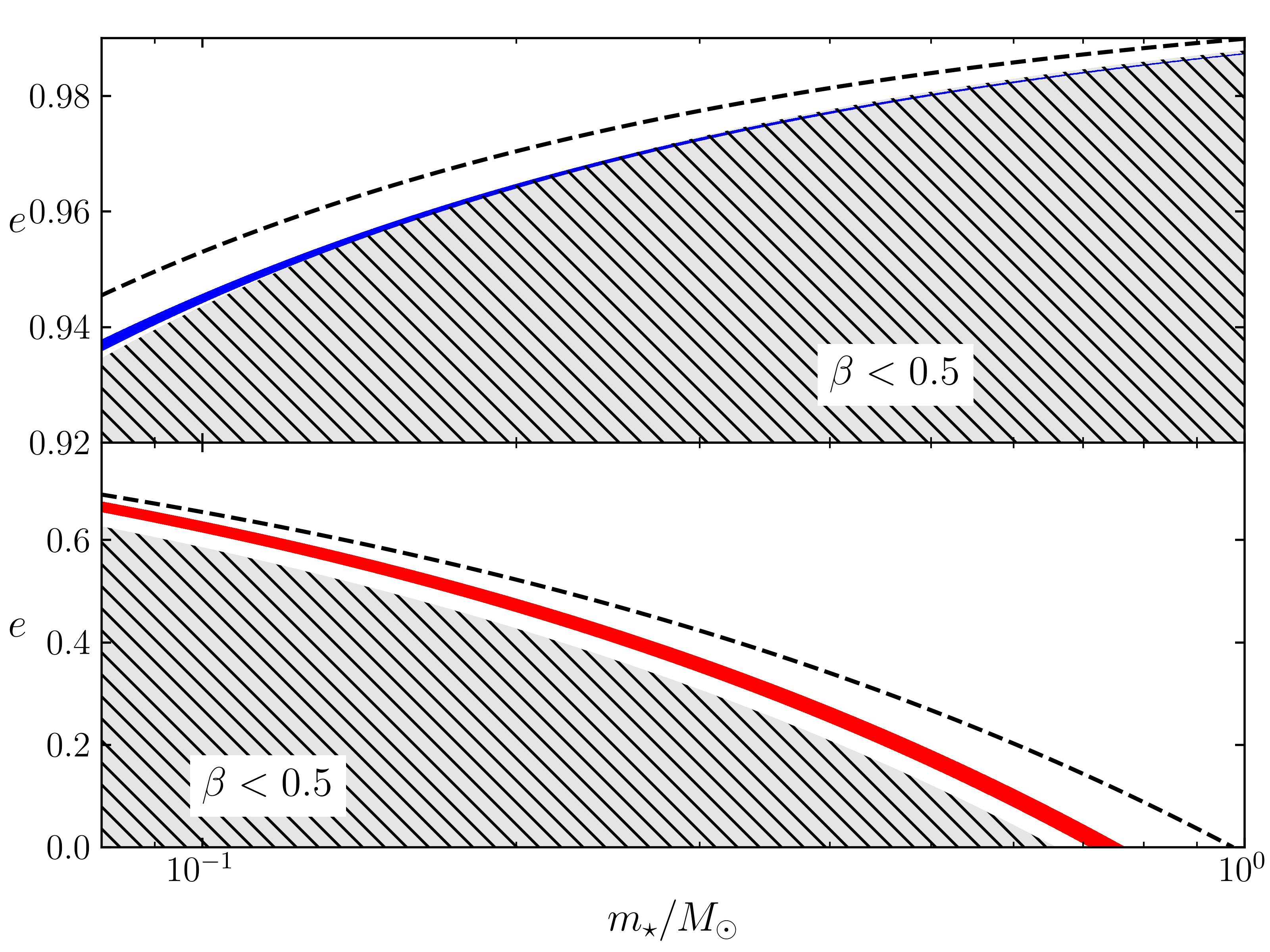}
    \caption{Orbital eccentricities and stellar mass values compatible with the BH mass and mass accretion rate obtained from the spectral fitting. Blue and red regions indicate WD and MS star cases respectively. The width of the contours covers the $5\sigma$ regions. The barred portions of parameter space are forbidden as they represent the regions where $\beta<0.5$ (no disruption occurs) and the dashed lines indicate the values for which $\beta=0.6$. \label{fig:beta}}
\end{figure}

Assuming we spotted the TDE close to its peak luminosity, we can use the mass accretion rate and BH mass obtained from the spectral fit to derive the penetration factor, orbital eccentricity and mass of the stripped star. To do so, we employed reasonable mass-radius relations for a white dwarf (WD) and for a main sequence star (MS),\footnote{We assumed the relations  $r_\textup{MS}/R_\odot = (m_\textup{MS}/M_\odot)^{0.8}$ and $r_\textup{WD}/R_\odot=0.01(m_\textup{WD}/M_\odot)^{-1/3}$ for a MS and a WD \rev{donor, respectively}.} coupled with the analytical formulas derived by \citet{gui13} for the $\gamma = 5/3$ polytrope case. Fig.\,\ref{fig:beta} shows the possible combination of star mass and orbital eccentricity compatible with the spectral estimates of mass accretion rate and BH mass. The blue and red contours represent the results obtained for a main sequence star and a white dwarf, respectively. The width of the contours corresponds to the $5\sigma$ region. 
The penetration factor $\beta$, defined as the ratio between pericentre distance and tidal radius, measures how deep the star dives into the potential well of the hole. As the tidal radius $r_\textup{t}\simeq r_\star(M_\textup{h}/m_\star)^{1/3}$, represents the distance at which the BH tidal forces balance the stellar self-gravity, for penetration factors $\beta<1$ the disruption is only partial. For values of $\beta\lesssim0.5$ the mass loss of the star stops \citep{gui13}: in Fig.\,\ref{fig:beta} grey bars cover these forbidden regions. 
As Fig.\,\ref{fig:beta} shows, only stars with masses below one solar mass are compatible with the spectral fitting results. Furthermore, for a WD only highly eccentric orbits are permitted, with $0.94\lesssim e \lesssim 0.96$, while for MS stars only values of eccentricities $\lesssim0.7$ are permitted. In any case, for both WDs and MS stars, the passage around the BH is shallow: the dashed black line in Fig \ref{fig:beta} represents the values for which  the penetration factor is $\beta=0.6$.
As for the case of XMMU\,J122939.7+075333, the amount of mass accreted at every passage is very little: about $5.5\times10^{-10}$\,$M_\odot$. As pointed out by \citet{tie22}, with these low values of mass getting accreted per passage, one such event can last for very long times and indeed XMMU\,J122939.7+075333 has been active for more than two decades.
\rev{Finally, although the available observations are sparse and often provide only shallow upper limits, the outbursts of \src\ seem to recur on a time scale of $\approx$5 months,\revvv{ which is the time interval between the two closest outbursts and cannot be ruled out by non-detections.}}\\

\begin{figure}
	\includegraphics[width=\hsize]{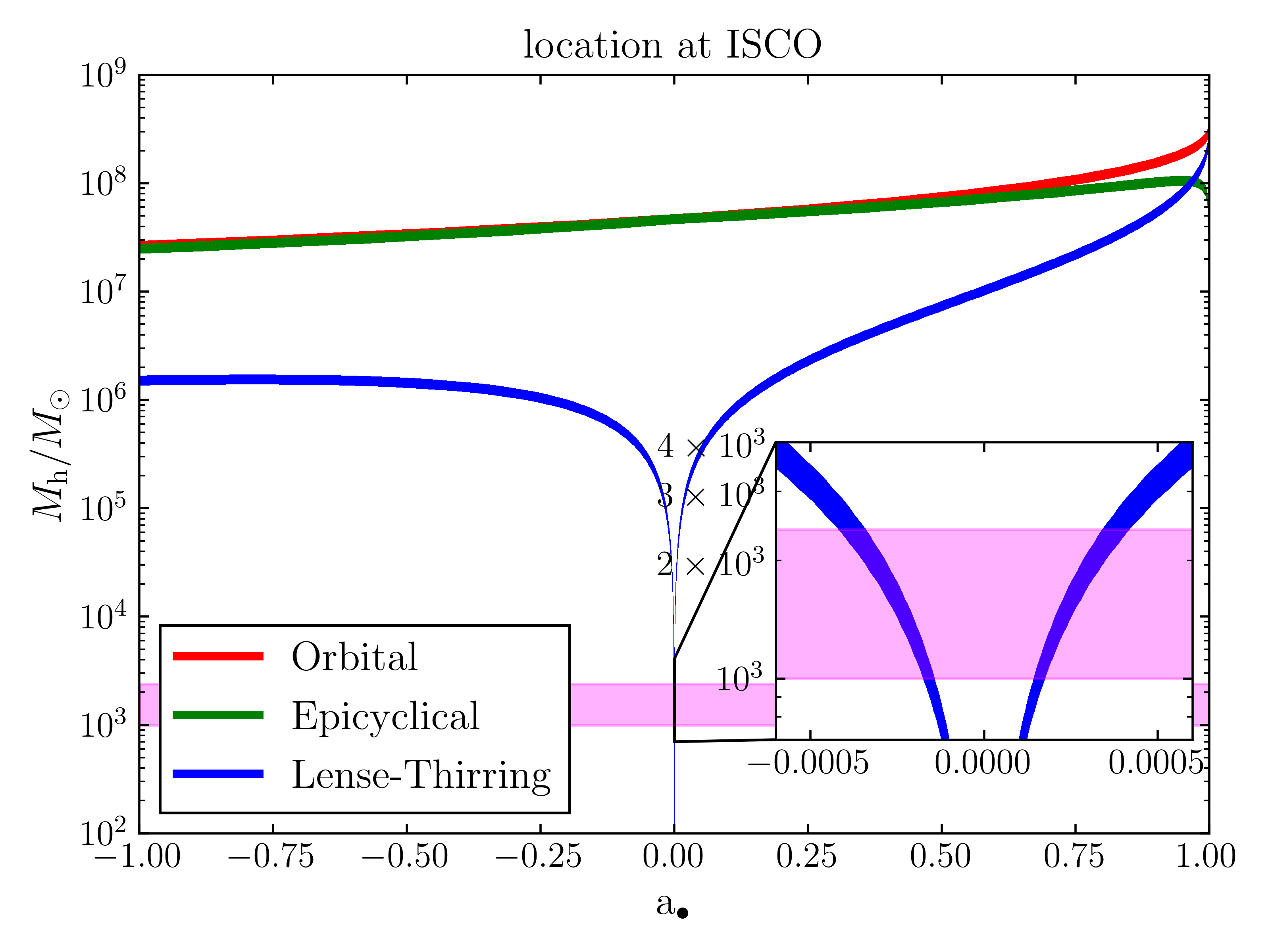}
    \caption{Contour plot in the BH mass-spin space of the three fundamental frequencies at ISCO for a 47 $\mu$Hz frequency. The shaded region indicates the BH mass range obtained from the spectral fit. The width of the lines reflects a $5\sigma$ range. \label{fig:freq}}
\end{figure}

The TDE scenario could also explain \src\ short-term variability: the observed periodicity could be linked to disc instability in a similar way to the one described by \citet{pas19} for the ``text-book TDE'' ASASSN-14li \citep{mil15}. To investigate this possibility, we can study the fundamental frequencies at the innermost stable circular orbit (ISCO) around the BH. These are three: the Keplerian orbital frequency $\nu_\phi$, the vertical epicyclic frequency $\nu_\theta$, and the Lense--Thirring one, given by the beating between these two, $\nu_\textup{LT}=\nu_\phi-\nu_\theta$. Analytical expressions for these frequencies were derived by \citet{kat90}:
\begin{equation}
\nu_\phi = \frac{c^3}{2\pi GM_\textup{h}}\left[\frac{1}{R_\textup{ISCO}^{3/2}+a_\bullet}\right],
\end{equation}
\begin{equation}
\nu_\theta =\nu_\phi\left[1-\frac{4a_\bullet}{R_\textup{ISCO}^{3/2}}+\frac{3a^2_\bullet}{R_\textup{ISCO}^2}\right]^{1/2}.
\end{equation}
Fig.\,\ref{fig:freq} shows the contours, in the BH mass-spin space, for the three described frequencies, Keplerian (in red),  vertical epicyclic (in green) and Lense--Thirring (in blue) for a $47\pm1$\,$\mu$Hz frequency (the $5.88\pm0.12$ h modulation). The width of the contours represents the $5\sigma$ region. The shaded region shows the BH mass range identified by the spectral fit, its width corresponds to a $\Delta C=1$. The only frequency compatible with the mass range identified by the spectral fitting is the Lense--Thirring one and only for a non-spinning BH: $|a_\bullet|\lesssim 4\times10^{-4}$.\\

The TDE scenario cannot account for every aspect of \src. The absorption feature in the X-ray spectrum, around 1\,keV is similar to ones often observed in super soft sources, both Galactic (nuclear burning WD, see e.g. \citealt{ebisawa01}) and extragalactic (ULXs, see e.g. \citealt{pinto21}), however, this feature is usually detected in sources with thick outflows rather than simple disc emission, and only at high values of mass accretion rates. Another weak point in this scenario is the involvement of a \rev{non-spinning} $10^3\,M_\odot$ IMBH. 
Albeit it has been shown that young dense clusters can nurture the seeding of IMBHs (\citealt{arcasedda2021,dicarlo21,rizzuto2021,rizzuto2022,gonzalez23}; Arca Sedda et al., in prep), it seems hard for YSCs to grow IMBHs above 1000\,$M_\odot$, unless the host YSC has large central densities \citep[e.g.][]{maliszewski2022} \revv{and the tightly constrained null value of the spin would require some fine tuning for the model to work}.
On the other hand, the TDE scenario can provide a satisfactory explanation for the X-ray spectral shape and both the long- and short-term behaviour of \src.
  
\begin{acknowledgements}
We thank the anonymous referee for the insightful comments and constructive report. 
This research is based on data and software provided by the NASA/GSFC’s High Energy Astrophysics Science Archive Research Center (HEASARC), the \cxo\ X-ray Center (CXC, operated for NASA by SAO), the ESA’s \xmm\ Science Archive (XSA), and on observations made with the {\em Southern African Large Telescope} (SALT) through the transient followup program 2018-2-LSP-001 (PI:DAHB). AS, PE, GLI, AT, and CP acknowledge financial support from the Italian Ministry for University and Research, through the grants 2017LJ39LM (UNIAM) and 2022Y2T94C (SEAWIND). DdM acknowledges financial support from INAF Mainstreams and AstroFund 2022 FANS projects grants. RS acknowledges grant number 12073029 from the National Natural Science Foundation of China (NSFC). MI is supported by the AASS PhD joint research program between the University of Rome "Sapienza" and the University of Rome "Tor Vergata", with the collaboration of the National Institute of Astrophysics (INAF). IMM and DAHB are supported by the South African NRF. AACS is supported by the German \emph{Deut\-sche For\-schungs\-ge\-mein\-schaft, DFG\/} in the form of an Emmy Noether Research Group -- Project-ID 445674056 (SA4064/1-1, PI Sander). AACS further acknowledges support from the Federal Ministry of Education and Research (BMBF) and the Baden-Württemberg Ministry of Science as part of the Excellence Strategy of the German Federal and State Governments. MAS acknowledges funding from the European Union’s Horizon 2020 research and innovation programme under the Marie Skłodowska-Curie grant agreement No. 101025436 (project GRACE-BH, PI: Manuel Arca Sedda). PE and AS thank M. Mapelli, L. Zampieri and G. Lodato for interesting and insightful discussion.
\end{acknowledgements}  

\bibliographystyle{aa} 
\bibliography{biblio} 

\appendix
\section{X-ray observations}
The journal of the X-ray observations is given in Tab. \ref{tab:cxo}.
\longtab[1]{
\begin{longtable}{cllcc}
\caption{Journal of the \cxo, \xmm\ and \swift/XRT observations used in this work. Errors are at $1\sigma$ and upper limits at $3\sigma$. The corresponding fluxes are shown in Fig.\,\ref{fig:longterm}.\label{tab:cxo}}\\
\hline
Instrument &Obs.ID & Date & Exposure & Rate \\   
& & & (ks) & (counts\,s$^{-1}$) \\
\hline
\endfirsthead
\caption{Continued.}\\
\hline
Instrument &Obs.ID & Date & Exposure & Rate \\   
& & & (ks) & (counts\,s$^{-1}$) \\
\hline
\endhead
\hline
\endfoot
\hline
\endlastfoot
\xmm(pn) & 0112800201 & 2000-12-26/27 &  30.0 &  $<2.64\times10^{-3}$ \\ 
\xmm(pn) & 0112800101 & 2001-01-01/02 & 40.0 & $<8.73\times10^{-4}$\\
\xmm(pn) & 0305860401 & 2005-05-22  & 28.1 & $<2.76\times10^{-3}$ \\
\xmm(MOS) & 0305860301 & 2005-11-25  & 36.0 & $<7.23\times10^{-4}$ \\
XRT & 00035866001 & 2006-09-05 & 0.8 & $<2.27\times10^{-2}$ \\
XRT & 00035866002 & 2006-11-01 & 0.6 & $<1.98\times10^{-2}$ \\
XRT & 00035866003 & 2006-11-07 & 1.8 & $<6.19\times10^{-3}$ \\
XRT & 00030848001 & 2006-12-27 & 2.0 & $<5.58\times10^{-3}$ \\
XRT & 00030848002 & 2006-12-29 & 9.9 & $<1.05\times10^{-3}$\\
XRT & 00030848003 & 2006-12-30 & 9.6 & $<1.07\times10^{-3}$\\
XRT & 00030848004 & 2006-12-31 & 9.4 & $<1.13\times10^{-3}$\\
XRT & 00030848005 & 2007-01-01 & 9.2 & $<1.60\times10^{-3}$\\
XRT & 00030848006 & 2007-01-02 & 7.7 & $<1.33\times10^{-3}$\\
XRT & 00030848007 & 2007-01-03 & 8.8 & $<1.21\times10^{-3}$\\
XRT & 00030848008 & 2007-01-04 & 2.3 & $<4.90\times10^{-3}$\\
XRT & 00030848009 & 2007-01-05 & 5.0 & $<2.65\times10^{-3}$\\
XRT & 00030848010 & 2007-01-06 & 8.7 & $<1.37\times10^{-3}$\\
XRT & 00030848011 & 2007-01-07 & 7.0 & $<1.76\times10^{-3}$\\
XRT & 00030848012 & 2007-01-08 & 3.8 & $<4.67\times10^{-3}$\\
XRT & 00035866004 & 2007-01-28/29 & 1.7 & $<6.65\times10^{-3}$ \\
XRT & 00031210001 & 2008-05-20 & 6.1 & $<2.14\times10^{-3}$ \\
XRT & 00031210002 & 2008-05-25 & 7.1 & $<1.85\times10^{-3}$\\
XRT & 00031210003 & 2008-06-04 & 6.9 & $<1.88\times10^{-3}$\\
XRT & 00031210004 & 2008-06-19 & 8.0 & $<2.77\times10^{-3}$\\
XRT & 00031210005 & 2008-08-14 & 2.1 & $<6.52\times10^{-3}$\\
XRT & 00031210006 & 2008-08-15 & 2.2 & $<6.00\times10^{-3}$\\
XRT & 00031210007 & 2008-08-26 & 4.7 & $<3.97\times10^{-3}$\\
XRT & 00031210008 & 2008-10-15 & 2.1 & $<5.28\times10^{-3}$\\
XRT & 00031210009 & 2008-10-16 & 1.6 & $<8.56\times10^{-3}$\\
XRT & 00031726001 & 2010-05-24/26 & 6.6 & $<2.01\times10^{-3}$ \\
XRT & 00031726002 & 2010-05-26 & 2.7 & $<5.19\times10^{-3}$\\
\xmm(pn) & 0656780401 & 2010-05-28  & 18.1 & $<4.42\times10^{-3}$ \\
XRT & 00031726003 & 2010-05-28 & 3.2 & $<4.27\times10^{-3}$\\
XRT & 00031726004 & 2010-05-30 & 3.2 & $<4.02\times10^{-3}$\\
XRT & 00031726005 & 2010-06-01 & 3.9 & $<4.68\times10^{-3}$\\
XRT & 00031726006 & 2010-06-03 & 3.0 & $<4.38\times10^{-3}$\\
XRT & 00031726007 & 2010-06-05 & 4.5 & $<2.89\times10^{-3}$\\
XRT & 00031726008 & 2010-06-07 & 4.1 & $<3.16\times10^{-3}$\\
XRT & 00031726009 & 2010-06-09 & 4.0 & $<3.28\times10^{-3}$\\
XRT & 00031726010 & 2010-06-11 & 3.8 & $<3.57\times10^{-3}$\\
XRT & 00031726011 & 2010-06-16 & 4.0 & $<4.49\times10^{-3}$\\
XRT & 00031726012 & 2010-06-17 & 4.2 & $<3.19\times10^{-3}$\\
XRT & 00031726013 & 2010-06-18 & 4.2 & $<3.23\times10^{-3}$\\
XRT & 00031726015 & 2010-06-21 & 2.2 & $<6.01\times10^{-3}$\\
XRT & 00031726016 & 2010-06-25 & 1.3 & $<1.07\times10^{-2}$\\
XRT & 00031726017 & 2010-07-01 & 4.0 & $<3.36\times10^{-3}$\\
\cxo & 12238 & 2010-09-24/25 & 63.0 & $<1.7\times10^{-4}$ \\
XRT & 00031726018 & 2010-09-26 & 2.3 & $<7.93\times10^{-3}$ \\
XRT & 00031726019 & 2010-09-27 & 3.5 & $<3.76\times10^{-3}$\\
XRT & 00031726020 & 2010-09-28 & 4.3 & $<3.08\times10^{-3}$\\
XRT & 00049834001 & 2013-11-27 & 0.5 & $<2.6\times10^{-2}$ \\
\cxo & 16028 & 2014-05-16/17 & 64.2 & $<9.3\times10^{-5}$ \\
\cxo & 16029 & 2014-11-17/18 & 61.3 & $(9.0\pm0.4)\times10^{-3}$ \\
XRT & 00049834002 & 2016-04-14 & 0.6 & $<2.38\times10^{-2}$ \\
XRT & 00049834003 & 2016-04-20 & 0.5 & $<2.83\times10^{-2}$\\
XRT & 00049834005 & 2016-04-24/25 & 2.9 & $<6.28\times10^{-3}$\\
\xmm(pn) & 0791010101 & 2016-12-17/18 &  102.6 &  $<1.72\times10^{-3}$ \\ 
\xmm(pn) & 0791010301 & 2016-12-19/20 & 47.5 & $<3.90\times10^{-3}$\\
XRT & 00049834006 & 2017-04-13 & 1.1 & $<1.26\times10^{-2}$ \\
XRT & 00049834007 & 2017-04-14/15 & 2.5 & $<5.23\times10^{-3}$\\
XRT & 00049834008 & 2017-04-16/7 & 1.9 & $<7.25\times10^{-3}$\\
XRT & 00049834009 & 2017-04-21 & 0.5 & $<2.63\times10^{-2}$\\
XRT & 00049834010 & 2017-04-24 & 5.2 & $<2.51\times10^{-3}$\\
XRT & 00049834012 & 2017-07-02 & 1.3 & $<8.76\times10^{-3}$ \\
XRT & 00049834013 & 2017-07-06 & 1.9 & $<7.02\times10^{-3}$\\
XRT & 00049834014 & 2017-07-12 & 1.8 & $<7.48\times10^{-3}$\\
XRT & 00049834015 & 2018-01-25 & 5.0 & $<2.25\times10^{-3}$\\
XRT & 00088651001 & 2018-01-31 & 5.1 & $<2.17\times10^{-3}$\\
XRT & 00049834016 & 2018-02-07 & 4.6 & $<3.91\times10^{-3}$ \\
XRT & 00049834017 & 2018-02-14 & 4.8 & $<2.83\times10^{-3}$\\
XRT & 00049834018 & 2018-04-12 & 0.3 & $<4.42\times10^{-2}$ \\
XRT & 00049834019 & 2018-04-13/14 & 3.5 & $<3.87\times10^{-3}$ \\
XRT & 00049834020 & 2018-04-18 & 2.6 & $<5.22\times10^{-3}$ \\
XRT & 00049834021 & 2018-04-25 & 6.1 & $<2.19\times10^{-3}$ \\
XRT & 00049834022 & 2018-04-26 & 5.2 & $<2.55\times10^{-3}$ \\
XRT & 00049834023 & 2018-04-27 & 4.9 & $<3.70\times10^{-3}$ \\
XRT & 00049834024 & 2018-04-28 & 4.8 & $<2.75\times10^{-3}$ \\
XRT & 00049834025 & 2018-04-29 & 4.9 & $<3.66\times10^{-3}$ \\
XRT & 00049834026 & 2018-04-30 & 5.0 & $<2.70\times10^{-3}$ \\
XRT & 00049834027 & 2018-05-01 & 4.7 & $<3.71\times10^{-3}$ \\
XRT & 00049834028 & 2018-05-02 & 5.0 & $<3.58\times10^{-3}$ \\
XRT & 00049834029 & 2018-05-03 & 4.9 & $<2.83\times10^{-3}$ \\
XRT & 00049834030 & 2018-05-04 & 4.8 & $<2.71\times10^{-3}$ \\
XRT & 00049834031 & 2018-05-05 & 1.5 & $<9.21\times10^{-3}$ \\
XRT & 00049834032 & 2018-05-06 & 4.5 & $<5.00\times10^{-3}$ \\
XRT & 00049834033 & 2018-05-07 & 3.2 & $<4.25\times10^{-3}$ \\
XRT & 00049834034 & 2018-05-08 & 6.2 & $<3.52\times10^{-3}$ \\
XRT & 00049834035 & 2018-05-11 & 4.9 & $<3.72\times10^{-3}$ \\
XRT & 00049834036 & 2018-05-14 & 4.9 & $<2.78\times10^{-3}$ \\
XRT & 00049834037 & 2018-05-17 & 4.8 & $<2.76\times10^{-3}$\\
XRT & 00049834038 & 2018-05-20 & 4.7 & $<2.95\times10^{-3}$\\
XRT & 00049834039 & 2018-05-23 & 4.9 & $<3.58\times10^{-3}$\\
XRT & 00049834040 & 2018-05-26 & 4.8 & $<2.80\times10^{-3}$\\
XRT & 00049834041 & 2018-05-29 & 4.7 & $<2.79\times10^{-3}$\\
XRT & 00049834043 & 2018-06-01 & 4.9 & $<2.80\times10^{-3}$\\
XRT & 00049834044 & 2018-06-07 & 5.2 & $<2.55\times10^{-3}$\\
XRT & 00049834045 & 2018-06-10 & 4.7 & $<2.86\times10^{-3}$\\
XRT & 00049834046 & 2018-06-13 & 4.9 & $<2.68\times10^{-3}$\\
XRT & 00049834047 & 2018-06-16 & 4.7 & $<2.87\times10^{-3}$\\
XRT & 00049834048 & 2018-06-19 & 4.7 & $<2.85\times10^{-3}$\\
XRT & 00049834049 & 2018-06-28 & 3.8 & $<3.60\times10^{-3}$\\
XRT & 00049834050 & 2018-07-03 & 4.3 & $<3.20\times10^{-3}$\\
XRT & 00049834051 & 2018-07-08 & 4.4 & $<2.53\times10^{-3}$\\
XRT & 00049834052 & 2018-07-13 & 3.7 & $<4.83\times10^{-3}$\\
XRT & 00049834053 & 2018-07-18 & 3.6 & $<5.12\times10^{-3}$ \\
XRT & 00049834054 & 2018-07-23 & 4.3 & $<3.09\times10^{-3}$ \\
XRT & 00049834055 & 2018-07-28 & 3.8 & $<3.64\times10^{-3}$ \\
XRT & 00049834056 & 2018-08-02 & 3.9 & $<3.45\times10^{-3}$ \\
XRT & 00049834057 & 2018-08-07 & 4.0 & $<3.38\times10^{-3}$ \\
XRT & 00049834058 & 2018-08-12 & 4.0 & $<3.38\times10^{-3}$ \\
XRT & 00049834059 & 2018-08-29 & 3.6 & $(1.9\pm1.0)\times10^{-3}$\\
XRT & 00049834060 & 2018-09-05 & 3.8 & $<3.52\times10^{-3}$ \\
XRT & 00049834061 & 2018-09-12 & 4.5 & $<3.02\times10^{-3}$ \\
XRT & 00049834063 & 2018-09-23 & 4.8 & $<2.73\times10^{-3}$ \\
XRT & 00049834064 & 2018-09-26 & 4.6 & $<2.87\times10^{-3}$ \\
XRT & 00049834065 & 2018-10-03 & 4.7 & $<2.76\times10^{-3}$ \\
XRT & 00088810002 & 2018-10-07 & 4.9 & $<2.66\times10^{-3}$ \\
XRT & 00049834066 & 2018-10-10 & 2.1 & $<6.39\times10^{-3}$ \\
XRT & 00049834067 & 2018-10-14 & 2.9 & $<4.40\times10^{-3}$ \\
XRT & 00049834068 & 2018-10-17 & 4.7 & $<2.27\times10^{-3}$ \\
XRT & 00049834070 & 2018-11-18 & 1.0 & $<1.17\times10^{-2}$ \\
XRT & 00049834071 & 2018-11-25 & 0.4 & $<3.41\times10^{-2}$ \\
XRT & 00049834072 & 2018-11-28 & 0.6 & $<2.31\times10^{-2}$ \\
XRT & 00049834073 & 2018-12-02 & 1.0 & $<1.43\times10^{-2}$ \\
XRT & 00049834074 & 2018-12-09 & 1.1 & $<1.18\times10^{-2}$ \\
XRT & 00049834075 & 2018-12-16 & 1.0 & $<1.41\times10^{-2}$ \\
XRT & 00049834076 & 2018-12-23 & 0.6 & $<1.79\times10^{-2}$ \\
XRT & 00049834077 & 2018-12-30 & 0.9 & $<1.46\times10^{-2}$ \\
XRT & 00049834078 & 2019-01-06 & 0.3 & $<4.21\times10^{-2}$ \\
XRT & 00049834080 & 2019-01-13 & 2.0 & $<6.83\times10^{-3}$ \\
XRT & 00049834081 & 2019-01-20 & 2.0 & $(3.4\pm1.8)\times10^{-3}$ \\
XRT & 00049834082 & 2019-01-27 & 1.8 & $<1.02\times10^{-2}$ \\
XRT & 00049834083 & 2019-02-03 & 1.8 & $<7.68\times10^{-3}$ \\
XRT & 00049834084 & 2019-02-10 & 2.0 & $<6.95\times10^{-3}$ \\
XRT & 00049834085 & 2019-04-13 & 2.1 & $<6.56\times10^{-3}$ \\
XRT & 00049834086 & 2019-04-20 & 2.5 & $<5.37\times10^{-3}$ \\
XRT & 00049834087 & 2019-04-27 & 0.4 & $<3.73\times10^{-2}$ \\
XRT & 00049834088 & 2019-05-04 & 2.3 & $<5.81\times10^{-3}$ \\
XRT & 00049834089 & 2019-05-11 & 2.1 & $<7.56\times10^{-3}$ \\
XRT & 00031726021 & 2019-09-20 & 1.9 & $<7.16\times10^{-3}$ \\
XRT & 00012017001 & 2019-09-26 & 2.5 & $<5.45\times10^{-3}$ \\
XRT & 00012017002 & 2019-09-26 & 3.0 & $<4.48\times10^{-3}$ \\
XRT & 00012017004 & 2019-09-27 & 3.1 & $<4.43\times10^{-3}$ \\
XRT & 00012160001 & 2019-11-11 & 0.8 & $<1.75\times10^{-2}$ \\
XRT & 00012160002 & 2019-11-14 & 0.8 & $<2.49\times10^{-2}$ \\
XRT & 00012160003 & 2019-11-16 & 1.2 & $<1.94\times10^{-2}$ \\
XRT & 00095672001 & 2020-04-16  & 3.7 & $(2.8\pm1.3)\times10^{-3}$ \\
\cxo & 22375 & 2020-04-26/27 & 47.4 & $(1.0\pm0.2)\times10^{-3}$ \\
XRT & 00095672002 & 2020-04-30  & 3.1 &  $<4.36\times10^{-3}$ \\
XRT & 00095672003 & 2020-05-14 & 3.8 & $<3.52\times10^{-3}$ \\
XRT & 00095672004 & 2020-05-28 & 3.3 & $<4.01\times10^{-3}$ \\
XRT & 00095672005 & 2020-06-11 & 3.6 & $<3.73\times10^{-3}$ \\
XRT & 00095672006 & 2020-06-25 & 3.3 & $<3.53\times10^{-3}$ \\
XRT & 00095672007 & 2020-07-09 & 3.7 & $<3.06\times10^{-3}$ \\
XRT & 00095672008 & 2020-07-23 & 3.8 & $<3.71\times10^{-3}$ \\
XRT & 00095672009 & 2020-08-06 & 2.9 & $<4.67\times10^{-3}$ \\
XRT & 00095672010 & 2020-08-20 & 3.1 & $<5.98\times10^{-3}$ \\
XRT & 00095672011 & 2020-09-03 & 3.8 & $<3.50\times10^{-3}$ \\
XRT & 00095672012 & 2020-09-17 & 3.9 & $(5.1\pm1.7)\times10^{-3}$ \\
XRT & 00095672013 & 2020-10-01 & 3.2 & $<5.79\times10^{-3}$ \\
XRT & 00095672014 & 2020-10-15 & 3.1 & $<4.38\times10^{-3}$ \\
XRT & 00095672015 & 2020-10-29 & 3.9 & $<3.40\times10^{-3}$ \\
XRT & 00095672016 & 2020-11-12 & 3.9 & $<2.79\times10^{-3}$ \\
XRT & 00095672017 & 2020-11-26 & 3.9 & $<2.81\times10^{-3}$\\
XRT & 00095672018 & 2020-12-10 & 2.3 & $<5.54\times10^{-3}$\\
XRT & 00095672019 & 2020-12-15 & 1.6 & $<1.08\times10^{-2}$ \\
XRT & 00095672020 & 2020-12-24 & 3.8 & $<5.87\times10^{-3}$ \\
XRT & 00095672021 & 2021-01-07 & 3.7 & $<2.92\times10^{-3}$ \\
XRT & 00095672022 & 2021-01-21 & 3.5 & $<3.14\times10^{-3}$ \\
XRT & 00095672023 & 2021-02-04 & 3.9 & $<3.38\times10^{-3}$ \\
XRT & 00031726022 & 2021-10-23 & 1.4 & $<9.29\times10^{-3}$ \\
XRT & 00031726023 & 2021-11-23 & 2.1 & $<5.11\times10^{-3}$ \\
XRT & 00031726024 & 2021-12-22 & 0.7 & $<1.99\times10^{-2}$ \\
XRT & 00031726025 & 2021-12-28 & 1.8 & $<8.58\times10^{-3}$ \\
XRT & 00031726028 & 2022-04-13 & 4.7 & $<2.78\times10^{-3}$ \\
XRT & 00031726029 & 2022-10-01 & 2.1 & $<7.43\times10^{-3}$ \\
XRT & 00031726030 & 2022-10-18 & 0.2 & $<6.80\times10^{-2}$\\
XRT & 00031726031 & 2022-10-29 & 1.5 & $<7.51\times10^{-3}$\\
XRT & 00031726033 & 2022-11-26 & 1.5 & $<8.98\times10^{-3}$ \\
XRT & 00031726034 & 2022-12-10 & 1.6 & $<1.13\times10^{-2}$ \\
\hline                                   
\end{longtable}
\label{tab:non_int}
}

\section{SALT spectrum}
\begin{figure}
    \includegraphics[height=\hsize, angle=-90]{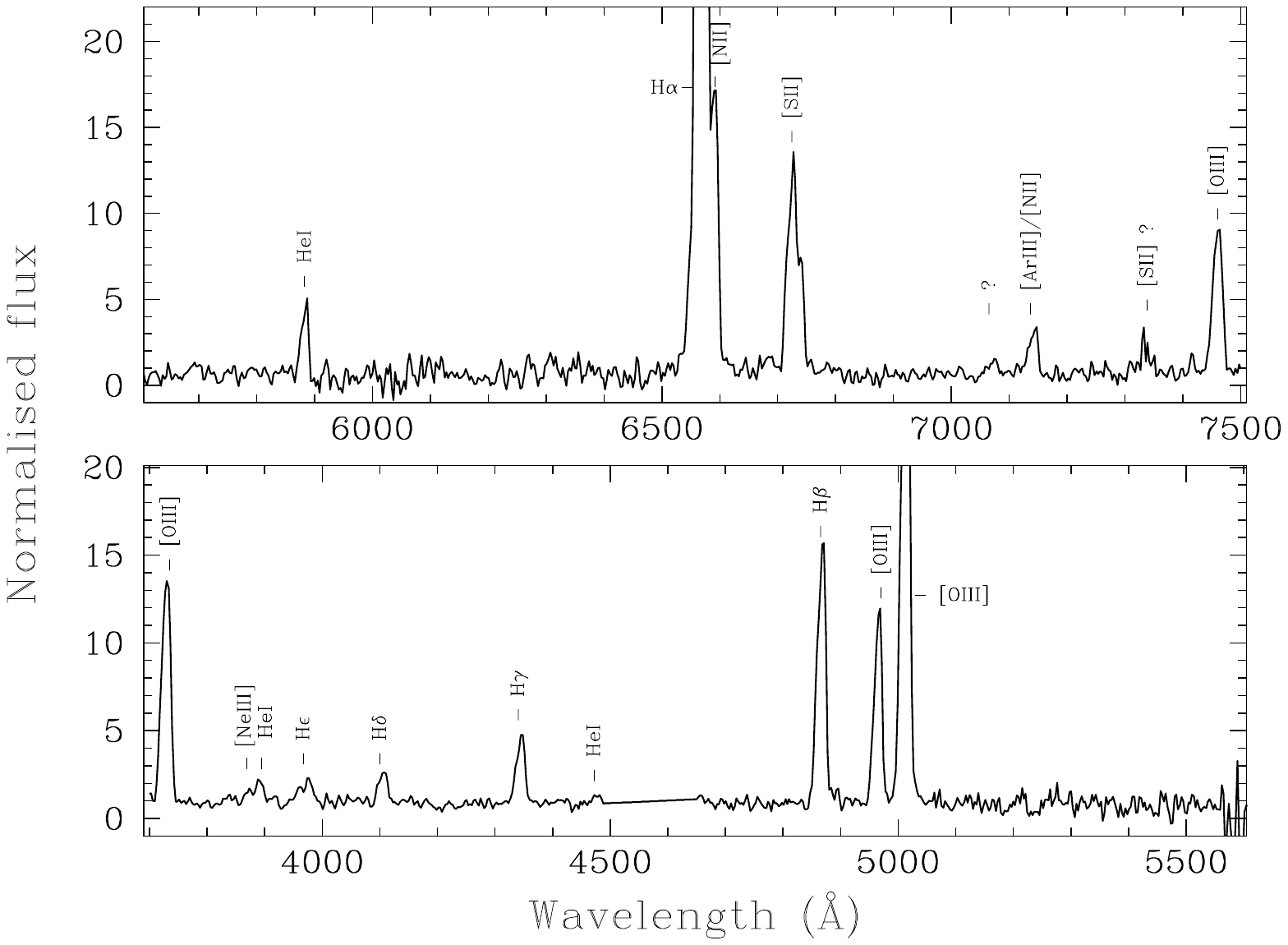}
    \caption{The SALT spectrum (PG0300) of \src. Note that the lines labelled with a question mark are better identified and resolved in the PG0900 spectrum (Fig.\,\ref{fig:salt}). \label{fig:salt_appendix}}
\end{figure}

The parameters of the emission lines in the optical SALT spectrum are reported in Tab. \ref{tab:salt}. The lower-resolution spectrum (PG0300) is shown in Fig. \ref{fig:salt_appendix}.

\begin{table*}
\caption{Parameters of emission lines in the two SALT spectra.\label{tab:salt}}
\centering
\begin{tabular}{lccccl} 
\hline\hline
Element        & $\lambda^1$    & FWHM           & EQW           &  V$_\textup{rad}^2$ &  Notes \\
               &   (\AA)         & (\AA)        & (\AA)       &   (km/s)    &   \\ 
\hline
\multicolumn{6}{c}{02 December 2019}\\
\hline
{[O\textsc{ii}]}    & $3729.1\pm0.2$     & $17.4\pm0.5$   & $332\pm13$    &  $121\pm17$ & blend of $3726-3729\,\AA$ \\
{[Ne\textsc{iii}]}  & $3872.3\pm0.1$     & $12.1\pm3.7$   & $9.7\pm3.7$   &  $252\pm12$ &  blended with He\textsc{i} $3888.64\,\AA$ \\
He\textsc{i}        & $3891.3\pm0.9$     & $14.0\pm2.2$   & $21.9\pm4.1$  &  $182\pm66$ &  blended with {[Ne\textsc{iii}]} $3868.75\,\AA$\\
{[O\textsc{iii}]}   & $3960.0\pm1.6$     & $11.7\pm3.8$   & $12.6\pm6.0$  & $-143\pm12$ &  uncertain identification\\
H$_\epsilon$       & $3976.8\pm1.1$     & $13.7\pm2.7$   & $24.0\pm8.2$  &  $483\pm83$ &  blended possibly with {[O\textsc{iii}]} $3961.59\,\AA$\\
H$_\delta$ & $4105.8\pm0.6$     & $16.6\pm1.5$   & $39.5\pm4.8$  &  $275\pm47$ & \\
H$_\gamma$ & $4344.8\pm0.3$     & $17.1\pm0.6$   & $99.6\pm4.9$  &  $273\pm18$ & \\
He\textsc{i}        & $4477.03\pm0.01$   & $36.74\pm0.01$ & $5.0\pm2.0$   &  $349.37\pm0.01$ &  double peaked, manually measured\\
H$_\beta$  & $4867.0\pm0.3$     & $16.7\pm0.7$   & $330\pm17$    &  $328\pm16$ &   with additional red-component\\
{[O\textsc{iii}]}   & $4965.5\pm0.3$     & $15.2\pm0.8$   & $240\pm18$    &  $375\pm19$ \\
{[O\textsc{iii}]}   & $5013.3\pm0.3$     & $16.0\pm0.9$   & $702\pm51$    &  $365\pm20$ & \\
He\textsc{i}        & $5883.4\pm0.6$     & $13.6\pm1.4$   & $113\pm15$    &  $376\pm2$9 & \\
H$_\alpha$ & $6569.7\pm0.3$     & $17.5\pm0.7$   & $1731\pm89$   &  $294\pm12$ &  blended with {[N\textsc{ii}]} $6548\,\AA$, and partially with $6584\,\AA$\\
{[N\textsc{ii}]}    & $6593.4\pm0.2$     & $16.1\pm0.6$   & $8.6\pm3.9$   &  $426.9\pm7.3$ &  blended with H$_\alpha$\\
{[S\textsc{ii}]}    & $6726.0\pm0.3$     & $20.5\pm0.8$   & $288\pm14$    &  $402\pm14$ &  blended with {[S\textsc{ii}]} $6731\,\AA$\\
{[N\textsc{ii}]}    & $7141.5\pm0.6$     & $18.3\pm1.4$   & $95.8\pm9.6$  &  $88\pm24$ &  uncertain identification or [Ar\textsc{iii}]$7136\,\AA$\\
{[Ar\textsc{iii}]}  &                    &                &               &  $208\pm24$ &  if {[Ar\textsc{iii}]}\\
{[O\textsc{iii}]}   & $7459.9\pm0.3$     & $18.3\pm0.7$   & $223\pm11$    &  $159\pm11$ & \\
\hline
\multicolumn{6}{c}{15 December 2019}\\
\hline
H$_\gamma$         & $4344.35\pm0.08$ & $5.4\pm0.2$ & $64.0\pm3.1$  & $242.5\pm5.7$ \\
H$_\beta$          & $4865.0\pm0.1$   & $5.6\pm0.3$ & $180\pm11$    & $206.0\pm6.5$ \\
{[O\textsc{iii}]}  & $4963.0\pm0.1$   & $5.5\pm0.4$ & $106.7\pm9.3$ & $225.1\pm9.1$ \\
{[O\textsc{iii}]}  & $5011.0\pm0.5$   & $5.4\pm0.1$ & $325.9\pm9.4$ & $224\pm29$ \\
He\textsc{i}       & $5879.9\pm0.1$   & $5.5\pm0.3$ & $92.3\pm7.4$  & $196.0\pm6.9$ \\
{[N\textsc{ii}]}   & $6553.4\pm0.8$   & $9.6\pm7.2$ & $77.3\pm7.5$  & $223\pm38$ \\  
H$_\alpha$         & $6567.73\pm0.08$ & $5.5\pm0.2$ & $1264\pm57$   & $200.9\pm3.6$  \\
{[N\textsc{ii}]}   & $6588.3\pm0.7$   & $5.8\pm1.6$ & $163\pm60$    & $200\pm30$ \\
He\textsc{i}       & $6683.6\pm0.2$   & $4.9\pm0.5$ & $15.0\pm2.0$  & $220.4\pm9.0$ \\
{[S\textsc{ii}]}   & $6721.3\pm0.1$   & $5.9\pm0.2$ & $181.5\pm9.6$ & $193.9\pm4.3$ \\  
{[S\textsc{ii}]}   & $6735.6\pm0.1$   & $6.1\pm0.3$ & $132.0\pm9.8$ & $190.4\pm6.2$ \\  
{[Ar\textsc{iii}]} & $7141.1\pm0.1$   & $4.9\pm0.4$ & $43.3\pm4.1$  & $192.9\pm6.9$ \\
\hline
\end{tabular}
\tablefoot{All parameters have been measured by Gaussian fits unless otherwise noted. $^1$ Line wavelength in \AA. $^2$ Radial velocity heliocentric corrected.}
\end{table*}

\end{document}